\documentclass[aps,prl,twocolumn,superscriptaddress]{revtex4-2}
\usepackage{cancel}
\usepackage[normalem]{ulem}
\usepackage{bm}
\usepackage{amsmath}
\usepackage{amsthm}
\usepackage{natbib}
\usepackage{amssymb}
\usepackage{graphicx}
\usepackage{esint}

\makeatletter
\usepackage{times}
\usepackage{comment}
\usepackage{graphicx}
\usepackage{feynmf}
\usepackage{tabularx}
\usepackage{amsmath}
\usepackage{amstext}
\usepackage{amssymb}
\usepackage{xfrac}
\usepackage[colorlinks,citecolor=blue]{hyperref}
\usepackage{graphicx}
\usepackage{amsmath}
\usepackage{amstext}
\usepackage{amssymb}
\usepackage{amsfonts}
\usepackage{longtable,booktabs}
\usepackage{hyperref}
\usepackage{url}
\usepackage{subfigure}
\usepackage{dsfont}
\usepackage{booktabs}
\usepackage{amsbsy}
\usepackage{dcolumn}
\usepackage{amsthm}
\usepackage{bm}
\usepackage{esint}
\usepackage{multirow}
\usepackage{hyperref}
\usepackage{cleveref}
\usepackage{mathrsfs}
\usepackage{amsfonts}
\usepackage{amsbsy}
\usepackage{dcolumn}
\usepackage{bm}
\usepackage{multirow}
\usepackage{color}
\usepackage{extarrows}
\usepackage{datetime}
\usepackage[super]{nth}
\hypersetup{
	colorlinks=magenta,
	linkcolor=blue,
	filecolor=magenta,
	urlcolor=magenta,
}

\newcommand{\bra}{\langle}
\newcommand{\ket}{\rangle}
\newcommand{\Tr}{\text{Tr}}

\newcommand{\tmmathbf}[1]{\ensuremath{\boldsymbol{#1}}}
\newcommand{\tmop}[1]{\ensuremath{\operatorname{#1}}}

\begin{document}
\title{Ward-Takahashi Identity and Gauge-Invariant Response Theory for Open Quantum Systems}
\author{Hongchao Li}
\affiliation{Department of Physics, The University of Tokyo, 7-3-1 Hongo, Tokyo 113-0033, Japan}
\email{lhc@cat.phys.s.u-tokyo.ac.jp}

\author{Xie-Hang Yu}
\affiliation{Max-Planck-Institut für Quantenoptik, Hans-Kopfermann-Straße 1, D-85748
	Garching, Germany}
\affiliation{Munich Center for Quantum Science and Technology, Schellingstraße
	4, 80799 München, Germany}

\author{Masaya Nakagawa}
\affiliation{Department of Physics, The University of Tokyo, 7-3-1 Hongo, Tokyo 113-0033,
	Japan}
\email{nakagawa@cat.phys.s.u-tokyo.ac.jp}

\author{Masahito Ueda}
\affiliation{Department of Physics, The University of Tokyo, 7-3-1 Hongo, Tokyo 113-0033,
	Japan}
\affiliation{RIKEN Center for Emergent Matter Science (CEMS), Wako, Saitama 351-0198,
	Japan}
\affiliation{Institute for Physics of Intelligence, The University of Tokyo, 7-3-1
	Hongo, Tokyo 113-0033, Japan}
\email{ueda@cat.phys.s.u-tokyo.ac.jp}

\date{\today}
\begin{abstract}
We derive the Ward-Takahashi identity and establish the gauge-invariant response theory for open quantum systems described by Lindbladians to show that particle-number conservation is not necessary to satisfy gauge invariance. We construct an observable which can be used to test the gauge invariance in the absence of particle-number conservation. We derive the low-energy collective modes that emerge as a consequence of gauge invariance in open quantum systems, and find that two-body loss induces diffusive modes in dissipative Bardeen-Cooper-Schrieffer (BCS) superconductivity. Possible experimental situations for testing gauge invariance in open quantum systems are also discussed.
\end{abstract}
\maketitle
\emph{Introduction.---} 
Spontaneous symmetry breaking is a key mechanism of second-order phase transitions in many-body systems~\cite{Leggett2006,RevModPhys.47.331,Coleman_2015,David_Symmetry,Ueda2010,Kibble_review,Lieu2020}. A prime example is superconductivity associated with U(1) symmetry breaking~\cite{Coleman_2015,Schmitt2015,RevModPhys.29.205,RevModPhys.47.331,Ueda2010,BCS1957,RevModPhys.71.463,Pethick_Smith_2008}. The Bardeen-Cooper-Schrieffer (BCS) mean-field Hamiltonian~\cite{BCS1957} breaks the U(1) symmetry, which leads to two unphysical consequences: violation of particle-number conservation and superposition of states with different particle numbers in the ground state. As a result, the response current in the BCS theory is gauge-dependent ~\cite{BCS1957,Anderson1958_SC}, which is problematic since the physical current should be independent of the gauge choice of the electromagnetic (EM) field. This difficulty raised a fundamental question of how to construct a gauge-invariant response theory of superconductivity~\cite{London1948,London1935,Anderson1958_SC,Bardeen1951}. Nambu resolved this question by invoking the Ward-Takahashi identity~\cite{Nambu1960,Ward1950,Takahashi1957} which is based on the U(1) symmetry of the Hamiltonian. An important consequence of gauge invariance is the emergence of a low-energy gapless collective mode, known as the Nambu-Goldstone (NG) mode~\cite{Nambu1960,Goldstone_theorem,Goldstone1962}.

Although the gauge invariance is discussed mostly for closed quantum systems, real quantum systems inevitably undergo dissipation as in ultracold atomic and molecular systems and quantum materials coupled to environments ~\cite{Ni2008,Ni2010,Ospelkaus2010,Liu2020,Bause2023,Bause2021,Sebastian2016,XingYan2022,Yoshida2018,Helfrich2010,Schindewolf2022,Theory_Open2007,Kamenev_2011,Aoki2014,Qinghong2021,Tsuji2009}, where the systems are open and the particle number is often not conserved. Dissipation can induce superfluidity~\cite{Hongchao2023,Yamamoto2019,Yamamoto2021,Hongchao2024,Han2009} and phase transitions~\cite{Diehl2010,Sieberer_2016,Nakagawa2021} unique to open systems, and enhance quantum transport~\cite{Yamamoto2020Re,Damanet2019,Hongchao2025}. A crucial problem is thus how to construct a gauge-invariant response theory in open quantum systems without particle-number conservation. Notably, U(1) symmetry in open quantum systems does not necessarily imply particle-number conservation~\cite{Albert2014}, making the criterion of gauge invariance nontrivial. Here, it is instructive to revisit the two unphysical features in the BCS theory: non-conservation of the particle number and superposition of different particle-number sectors. In fact, they are equivalent for pure states in closed systems, but inequivalent in open quantum systems. We find that the most fundamental requirement for gauge invariance is the absence of superposition of different particle-number sectors, rather than particle-number conservation.

%Such Lindblad dynamics exhibits weak U(1) symmetry, which leads to the superselection rule in the density matrix, in contrast to the strong U(1) symmetry that respects the particle-number conservation. These two types of symmetry are equivalent in closed quantum systems but inequivalent in open quantum systems. \textcolor{red}{A crucial question is if we can construct a gauge-invariant theory in open quantum systems under the Lindbladian dynamics with weak U(1) symmetry instead of strong U(1) symmetry}.\\

%Though the nonunitary BCS theory has been studied~\cite{Hongchao2023,Yamamoto2019,Yamamoto2021,GMazza2023}, how to couple the EM field to a dissipative superconducting system and construct a gauge-invariant response theory remains elusive to us. It is of fundamental importance to investigate the exsitence of gauge invariance in open quantum systems without particle conservation.

In this Letter, we develop a gauge-invariant linear response theory for open quantum systems described by the local Lindblad dynamics. In particular, on the basis of the weak U(1) symmetry of the Lindbladian that commutes with the generator of phase rotation~\cite{Albert2014}, we derive the Ward-Takahashi identity for an open quantum system and apply it to show the gauge invariance even in the absence of particle-number conservation. We use the derived Ward-Takahashi identity to construct a general linear response theory for weak U(1) symmetric Lindbladian. Our result only relies on the symmetry of Lindbladians regardless of the specfic form of interactions and Lindblad operators. 
%The underlying physics is that the coherence part in the density matrix on the particle-number basis remains invariant under the weak U(1) symmetry, which is also supported by $f$-sum rule~\cite{Hongchao2024}.

Since there is no particle-number conservation in the Lindblad dynamics with weak U(1) symmetry, the particle number does not serve as an indicator of the gauge invariance. Instead, we employ the Ward-Takahashi identity to construct an observable $O_N(t)$ [Eq.~\eqref{eq: OTOC}] that can be used to test gauge invariance in open quantum systems and can be measured with the state-of-the-art techniques in cold-atom experiments~\cite{Daley2012,Islam2015,Pichler2016}. Furthermore, we derive the low-energy excitation spectrum as an important consequence of weak U(1) symmetry breaking for dissipative BCS superconductivity under two-body loss and find that dissipation induces a diffusive propagation of the NG mode. Compared with Refs. \cite{Minami2018,Hidaka2020} which examine steady states, our study concerns weak U(1) symmetry breaking in the Lindbladian dynamics. Lastly, we discuss possible experimental situations to test our results.

\emph{Ward-Takahashi identity.--- } We consider interacting fermions described by the Hamiltonian
\begin{align}
	H=&\sum_{\sigma}\int d^d\bm{r}\left(\frac{1}{2m}\nabla c_{\bm{r}\sigma}^{\dagger}\nabla c_{\bm{r}\sigma}+V_{\mathrm{int}}\right),\label{eq:hamiltonians}
\end{align}
where $d$ is the spatial dimension, $m$ is the mass of a single fermion, $c_{\bm{r}\sigma}^{(\dagger)}$ is the annihilation (creation) operator of a fermion with spin $\sigma$ at position $\bm{r}$, $V_{\mathrm{int}}$ represents the interaction between fermions which is a function involving only local density operators, and we set $\hbar=1$. The Hamiltonian satisfies the global U(1) symmetry, i.e., the Hamiltonian remains invariant under the transformation $c_{\bm{r}\sigma}\to c_{\bm{r}\sigma}e^{i\theta}$ where $\theta$ is a global constant. We couple the EM field to the fermionic system with the substitution $\partial_{\mu}\to\partial_{\mu}-i A_{\mu}$ where $\mu=0,\cdots,d$ indicates the space-time component and we set the coupling constant $e$ to unity. The correlation function is defined by
\begin{align}
	C (x_1, x_2) &:= \frac{i}{Z} \int D [\Psi, \bar{\Psi}]
	\Psi (x_1) \bar{\Psi} (x_2) e^{i S [A_{\mu}, \Psi,\bar{\Psi}]} \nonumber\\
	&=:i\langle\Psi (x_1) \bar{\Psi} (x_2)\rangle,
\end{align}
where $\Psi :=(c_{\uparrow}(\bm{r},t),\bar{c}_{\downarrow}(\bm{r},t))^T$ and $ \bar{\Psi} :=(\bar{c}_{\uparrow}(\bm{r},t),c_{\downarrow}(\bm{r},t))$ are the Nambu spinors, $S$ is an action, and $Z$ is the normalization factor. In the case of closed quantum systems, we have the Ward-Takahashi identity as
\begin{align}\label{eq: Ward2}
	&[\delta (x - x_1) \tau_3C(x_1,x_2) - \delta (x - x_2) C(x_1,x_2)\tau_3] \nonumber\\
	&=i\langle \Psi (x_1) \bar{\Psi} (x_2)
	\partial_{\mu} J^{\mu}(x) \rangle,
\end{align}
where $J^{\mu}(x)$ is the current operator and we use $\tau_{1,2,3}$ to represent the Pauli-$x,y,z$ matrices. Upon Fourier transformation, Eq. \eqref{eq: Ward2} reproduces the result in Ref. \cite{Nambu1960}. 

We now consider the case of open quantum systems described by the Lindblad equation \citep{Theory_Open2007}
\begin{equation}
\frac{d\rho}{dt}=\mathcal{L}[\rho]=-i[H,\rho]-\frac{\gamma}{2}\int d\bm{r}(\{L_{\bm{r}}^{\dagger}L_{\bm{r}},\rho\}-2L_{\bm{r}}\rho L_{\bm{r}}^{\dagger}),\label{eq: Lindblad}
\end{equation}
where $\rho$ is the density matrix of the system, $\mathcal{L}$ is the Lindbladian, and the Lindblad operator $L_{\bm{r}}$ describes local dissipation at position $\bm{r}$ with a dissipation rate $\gamma>0$. We first construct the gauge-invariant response theory by using the Schwinger-Keldysh field theory~\citep{Sieberer_2016}. We consider the path-integral representation of Eq.~(\ref{eq: Lindblad}) on the Schwinger-Keldysh contour with the action given by \citep{Sieberer_2016}
\begin{align}\label{eq:Keldysh_action}
	S= & \int_{-\infty}^{\infty}dt\biggl[\int d\bm{r}(i \bar{\Psi}_+	\partial_t \Psi_+ - i \bar{\Psi}_- \partial_t \Psi_-) - H_++H_{-}\nonumber\\
	&+\frac{i\gamma}{2}\int d\bm{r}(\bar{L}_{\tmmathbf{r}+}L_{\tmmathbf{r}+}+\bar{L}_{\tmmathbf{r}-}L_{\tmmathbf{r}-}-2L_{\tmmathbf{r}+}\bar{L}_{\tmmathbf{r}-})\biggr],
\end{align}
where the subscripts $+$ and $-$ denote the forward and
backward contours, $\Psi_{\alpha} :=(c_{\alpha\uparrow}(\bm{r},t),\bar{c}_{\alpha\downarrow}(\bm{r},t))^T$ and $ \bar{\Psi}_{\alpha} :=(\bar{c}_{\alpha\uparrow}(\bm{r},t),c_{\alpha\downarrow}(\bm{r},t))$ represent the Nambu spinors on the contour $\alpha=\pm$, and $c_{\alpha\sigma}(\bm{r},t)$ with $\sigma=\uparrow,\downarrow$ describes a fermionic field with spin $\sigma$. In the following we denote $x:=(t,\bm{r})$ for brevity. Here the Hamiltonian $H_\alpha$ and the Lindblad operators $L_{\bm{r}\alpha}(\bar{L}_{\bm{r}\alpha})$ are given by replacing $c_{\bm{r}\sigma}$ ($c_{\bm{r}\sigma}^\dag$) in Eq.~\eqref{eq:hamiltonians} with $c_{\alpha\sigma}(x)$ ($\bar{c}_{\alpha\sigma}(x)$). We require that the action satisfies the global weak U(1) symmetry~\cite{Albert2014,Buca_2012}, i.e., the invariance under $c_{\alpha\sigma}\to c_{\alpha\sigma}e^{i\theta}$, and that the dissipative part satisifes local weak U(1) symmetry, i.e., $	\bar{L}_{\tmmathbf{r}+}L_{\tmmathbf{r}+},\ \bar{L}_{\tmmathbf{r}-}L_{\tmmathbf{r}-}$, and $2L_{\tmmathbf{r}+}\bar{L}_{\tmmathbf{r}-}$ are invariant under $c_{\alpha\sigma}\to c_{\alpha\sigma}e^{i\theta(x)}$, where $\theta(x)$ is a function of space and time. An important example is two-body loss, where $L_{\tmmathbf{r}\alpha}=c_{\alpha\uparrow}(x)c_{\alpha\downarrow}(x)$~\cite{Yamamoto2019}. We couple the EM field to the fermionic system with the substitution $\partial_{\mu}\to\partial_{\mu}-i A_{\alpha\mu}$. The action coupled with the field $A_{\alpha\mu}$ satisifies the local weak U(1) symmetry under the transformations
\begin{equation}\label{eq: strong_U_1}
	\begin{aligned}
	&A_{\alpha\mu}\to A_{\alpha\mu}+\partial_{\mu}\theta(x),\ \Psi_{\alpha}\to e^{i\theta(x)\tau_3}\Psi_{\alpha},\\
	&\quad\quad\quad\quad\quad \bar{\Psi}_{\alpha}\to \bar{\Psi}_{\alpha}e^{-i\theta(x)\tau_3}.
	\end{aligned}
\end{equation}
\begin{figure}
	\includegraphics[width=0.9\columnwidth]{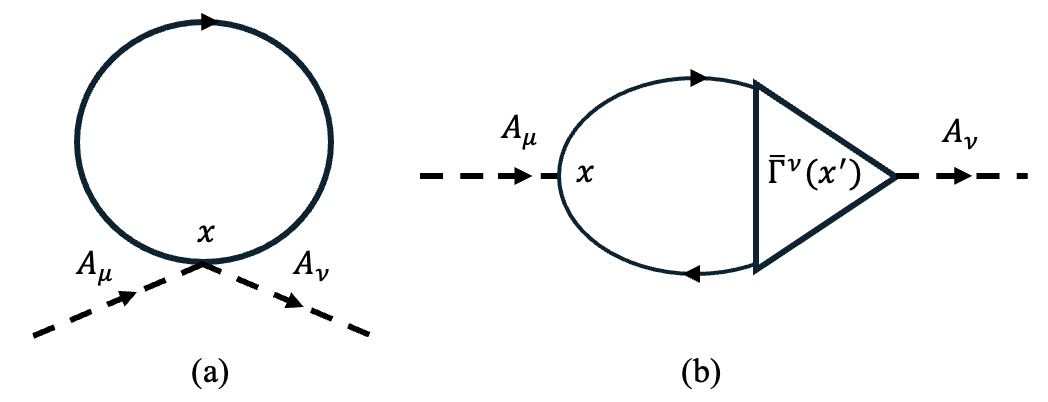}
	\caption{Feynman diagrams for linear response to an external EM field. Here the solid and dashed lines represent the fermion and photon propagators, respectively, and $\bar{\Gamma}^{\nu}$ in the second diagram represents the full vertex including interactions and dissipation and is defined from $\bar{J}_c^{\nu}(x)=\bar{\Psi}(x)\bar{\Gamma}^{\nu}(x)\Psi(x)$.}
	
	\label{fig:Feynman}
\end{figure}
Then we perform the local strong U(1) transformation to the correlation function $C_{\alpha\beta}:=i\langle\Psi_{\alpha} (x_1) \bar{\Psi}_{\beta} (x_2)\rangle$: 
\begin{equation}\label{eq: strong_U1_2}
	\begin{aligned}
	&A_{\alpha\mu}\to A_{\alpha\mu}+\alpha\partial_{\mu}\theta(x),\ \Psi_{\alpha}\to e^{i\alpha\theta(x)\tau_3}\Psi_{\alpha},\\
	&\quad\quad\quad\quad\quad \bar{\Psi}_{\alpha}\to \bar{\Psi}_{\alpha}e^{-i\alpha\theta(x)\tau_3}.
	\end{aligned}
\end{equation}
By taking the limit $\theta(x)\to0$ and requiring the terms linear in $\theta(x)$ to vanish, we obtain the following Ward-Takahashi identity in open quantum systems as
\begin{align}\label{eq: Ward}
	&\alpha[\delta (x - x_1) \tau_3C_{\alpha\alpha}(x_1,x_2) - \delta (x - x_2) C_{\alpha\alpha}(x_1,x_2)\tau_3] \nonumber\\
	&=i\langle \Psi_{\alpha} (x_1) \bar{\Psi}_{\alpha} (x_2)
	\partial_{\mu} \bar{J}_c^{\mu}(x) \rangle.
\end{align}
Importantly, the total current is defined as $\bar{J}_c^{\mu}:=J_c^{\mu}+J_d^{\mu}=J_{+}^{\mu}+J_{-}^{\mu}+J_d^{\mu}$ with
\begin{align}\label{eq: j_c}
	&\bm{J}_{\alpha}:=\frac{i}{2m}(\bar{\Psi}_{\alpha}\nabla\Psi_{\alpha}-\nabla\bar{\Psi}_{\alpha}\Psi_{\alpha}), J_{\alpha}^0:=\sum_{\sigma}n_{\alpha\sigma},\\ \label{eq: jd}
	& \nabla \cdot \tmmathbf{J}_d = i\gamma \left(
	\bar{L}_{\tmmathbf{r}-} \frac{\partial L_{\tmmathbf{r}+}}{\partial
		\theta(x)}\Big|_{\theta=0} + \frac{\partial \bar{L}_{\tmmathbf{r}-}}{\partial
		\theta(x)}\Big|_{\theta=0} L_{\tmmathbf{r}+} \right),\ J_d^0=0,
\end{align}
where $\partial B_{\bm{r}\alpha}/\partial \theta(x)|_{\theta=0}$ is the first-order derivative of an arbitrary field $B_{\bm{r}\alpha}$ under strong U(1) transformation~\eqref{eq: strong_U1_2}. Here $J_c^{\mu}$ is the kinetic current from the Hamiltonian and $J_d^{\mu}$ is the dissipative current from the dissipative part~\cite{PhysRevLett.93.160404}. A detailed derivation is shown in Supplemental Material~\cite{SupplementaryMaterial}. The dissipative current comes from the continuity equation given by
\begin{align}
	\frac{d \bra n_{\tmmathbf{r}}\ket}{d t} 	=-\nabla\cdot\bra\bm{j}_c+\bm{j}_d\ket,
\end{align}
where $n_{\bm{r}}$ is the fermion number density, $\bm{j}_c:=i(\Psi^{\dagger} \nabla \Psi - \nabla\Psi^{\dagger} \Psi)/2m$ and $\nabla \cdot \tmmathbf{j}_d = \frac{i \gamma}{2} \left(
L_{\tmmathbf{r}}^{\dagger} \frac{\partial L_{\tmmathbf{r}}}{\partial\theta_{\tmmathbf{r}}}|_{\theta=0} - \frac{\partial L_{\tmmathbf{r}}^{\dagger}}{\partial\theta_{\tmmathbf{r}}}|_{\theta=0} L_{\tmmathbf{r}} \right)$ with  $\partial B_{\bm{r}}/\partial \theta(x)|_{\theta=0}:=i[n_{\bm{r}},B_{\bm{r}}]$. It follows that $\bm{J}_c:=\bm{J}_++\bm{J}_-$ with Eq. \eqref{eq: j_c} is twice $\bm{j}_c$ and Eq. \eqref{eq: jd} is twice $\nabla\cdot\tmmathbf{j}_d$ if we place the operators $B_{\bm{r}}$ and $\frac{\partial B_{\bm{r}}^{\dagger}}{\partial\theta_{\tmmathbf{r}}}|_{\theta=0}$ on different contours in the path-integral representation where $B_{\bm{r}}=L_{\bm{r}}(L_{\bm{r}}^{\dagger})$. We note that $\bm{J}_d$ agrees with $\bm{j}_d$ if and only if the dissipative part of the Lindbladian satisfies weak U(1) symmetry. Physically, this fact is due to the decoupling of the diagonal and off-diagonal elements in the density matrix in the particle-number basis. Therefore, the full current $\bar{J}_c^{\mu}$ serves as the conserved response current satisfying $\partial_{\mu}\bar{J}_c^{\mu}=0$ in open quantum systems. In closed quantum systems, we take $\alpha=+$ in Eq. \eqref{eq: Ward} and recover the same result in Eq. \eqref{eq: Ward2}.

\emph{Gauge-invariant response theory.--- }We now prove the gauge invariance of the current in response to the EM field by applying the Ward-Takahashi identity. The perturbed current is given by
\begin{equation}
	\delta J_c^{\mu}(x)=-K^{\mu\nu}A_{\nu}(x),
\end{equation}
where the matrix $K^{\mu\nu}$ is composed of two Feynman diagrams shown in Fig. \ref{fig:Feynman}. Under the gauge transformation $A_{\mu}\to A_{\mu}+\partial_{\mu}\phi(x)$, one can directly show that the current changes $\delta J_1$ and $\delta J_2$ from the two Feynman diagrams by
\begin{equation}
	\delta J_1^{\mu}(x)=-\delta J_2^{\mu}(x)=-\frac{n(t)}{m}(1-\delta_{\mu0})\partial_{\mu}\phi(x),
\end{equation}
%\begin{align}
%	\delta J_1^{\mu}&=\delta_{\mu\nu}(1-\delta_{\mu0})\frac{1}{2m}\sum_{\alpha=\pm}\Tr[\tau_3\langle\Psi_{\alpha}(x)\bar{\Psi}(x)\rangle]\partial_{\nu}\phi\nonumber\\
%	&=-\frac{n(t)}{m}(1-\delta_{\mu0})\partial_{\mu}\phi(x),
%\end{align}
%\begin{align}
%	\delta J_2^{\mu} (x) &= \frac{1}{2} \sum_{\alpha} \int d^{d+1} x' \tmop{Tr}	\langle \gamma_{\alpha}^{\mu} \Psi_{\alpha} (x) \bar{J}_c^{\nu} (x')	\bar{\Psi}_{\alpha} (x) \rangle \partial_{\nu}' \phi (x')\nonumber\\
%	&=\frac{n(t)}{m}(1-\delta_{\mu0})\partial_{\mu}\phi(x),
%\end{align}
where
\begin{equation}
	n(t):= - \frac{1}{2} \int \frac{d^{d+1} k}{(2\pi)^{d+1}} \tmop{Tr} [\tau_3 (i G^T (k, t) - i
	G^{\tilde{T}} (k, t))],
\end{equation}
with $G^{T(\tilde{T})}(k,t)$ being the time-ordered (anti-time-ordered) Green's function defined as
\begin{equation}\label{eq: Green}
	\begin{aligned}
		&G^{T}\left(k,t\right)=\int d^{d+1}x e^{ik\cdot (x_1-x_2)} C_{++}(x_1,x_2), \\
		&G^{\tilde{T}}(k,t)=\int d^{d+1}x e^{ik\cdot (x_1-x_2)} C_{--}(x_1,x_2),
	\end{aligned}
\end{equation}
 where $t:=(t_1+t_2)/2$. The Green's functions \eqref{eq: Green} depend not only on the frequency but also on time due to the absence of time translational symmetry. The detailed derivation can be found in Supplemental Material~\cite{SupplementaryMaterial}. We can see that the change of the current vanishes $\delta J_c^{\mu}=\delta J_1^{\mu}+\delta J_2^{\mu}=0$. Hence, we have proved the gauge invariance of the current. Our results indicate that the gauge invariance can be maintained even without particle-number conservation. The key point of the proof is that the full vertex $\bar{\Gamma}^{\nu}$ in Fig. \ref{fig:Feynman} (b) takes into account both the conventional response current $J_c$ and the dissipative current $J_d$, unlike the case of closed quantum systems.

Since the current is gauge invariant, we take the Hamilton gauge $A_0=0$ and write the equation for the response current as a function of momentum and time as
\begin{equation}\label{eq: response}
	 \delta J_c^{\mu} (\tmmathbf{q}, t) = - \frac{n (t)}{m} (1 - \delta_{\mu 0})
	\left( \delta^{\mu j} - \frac{q^{\mu} q^j}{| \tmmathbf{q} |^2} \right) A_j
	(\tmmathbf{q}, t),
\end{equation}
where $j=1,\cdots,d$ is the space index. Thus, we have obtained an explicit form of dynamical gauge-invariant response current, which is a consequence of weak U(1) symmetry regardless of the type of dissipation. The current also satisfies the conservation law: $\partial_{\mu}\delta J_c^{\mu}(x)=0$. We note that the gauge invariance can also be considered as a consequence of the generalized $f$-sum rule derived in Ref. \cite{Hongchao2024}. Under the longitudinal gauge $\nabla\times\bm{A}=0,A_0=0$, we apply the generalized $f$-sum rule to show that $\bm{q}\cdot \delta\bm{J}_c(\bm{q},t)=0$, indicating that the response current is independent of the gauge choice of the EM field (see Supplemental Material~\cite{SupplementaryMaterial} for details).

The coefficent $n(t)$ in Eq. \eqref{eq: response} is determined by Green's functions and depends on the specific form of Hamiltonians and Lindblad operators. Here we take the dissipative BCS superconductivity as an example~\cite{Yamamoto2021}, with the Lindblad equation given by Eq. \eqref{eq: Lindblad} where the Hamiltonian is the BCS Hamiltonian $H=\sum_{\bm{k},\sigma}\varepsilon_{\bm{k}}c^{\dagger}_{\bm{k}\sigma}c_{\bm{k}\sigma}-U\int d\bm{r} c^{\dagger}_{\bm{r}\uparrow}c^{\dagger}_{\bm{r}\downarrow}c_{\bm{r}\downarrow}c_{\bm{r}\uparrow}$ with $\varepsilon_{\bm{k}}$ being the kinetic energy and $U>0$ being the interaction strength. The Lindblad operator describes on-site two-body loss, i.e., $L_{\bm{r}}=c_{\bm{r}\downarrow}c_{\bm{r}\uparrow}$. By employing the mean-field approximation, the dynamics of the density matrix obeys~\cite{Yamamoto2021}
 \begin{equation}
 	\frac{d\rho}{dt}=-i[H_{\text{BCS}},\rho],
 \end{equation}
 where $H_{\text{BCS}}$ is the mean-field Hamiltonian given by
 \begin{equation}
 	H_{\text{BCS}}=\sum_{\bm{k}}\Psi_{\bm{k}}^{\dagger}\begin{pmatrix} \varepsilon_{\bm{k}} & \Delta\\
 		\Delta^{*} & -\varepsilon_{\bm{k}}
 	\end{pmatrix}\Psi_{\bm{k}},\ \Delta=-\frac{U_c}{V}\sum_{\bm{k}}\langle c_{-\bm{k}\downarrow}c_{\bm{k}\uparrow}\rangle.
 \end{equation} 
  Here $U_c=U+i\gamma/2$ is complex and $V$ is the volume. When the dissipation rate is small, we make the quasi-steady-state approximation under which the particle number is nearly invariant within a long time period, and obtain $n(t)=N(t)/V$, where $N(t)$ is the number of fermions at time $t$~\cite{SupplementaryMaterial}. Hence, $n(t)$ represents the number density of fermions, whose dynamics has been studied in Refs.~\cite{Yamamoto2021,Mazza2023}. 
\begin{table*}[t]
	\centering
	\begin{tabular}{|c|c|c|c|}
		\hline
		symmetry & particle number $N(t)$ & $O_N(t)$ (defined in Eq. \eqref{eq: OTOC}) & gauge invariance  \\ \hline
		strong U(1) symmetry & conserved  & conserved & satisfied \\ \hline
		weak U(1) symmetry   & not conserved & conserved & satisfied \\ \hline
		no symmetry          & not conserved  & not conserved & not satisfied \\ \hline
	\end{tabular}
	\caption{Relations between symmetries, observables and gauge invariance.}
	\label{table:relation}
\end{table*}

\emph{Criterion for gauge invariance.--- } Equation \eqref{eq: response} shows the gauge-invariant response current based on weak U(1) symmetry. When the Lindbladian violates weak U(1) symmetry, we can see that the perturbed current is not invariant under the gauge transformation as
\begin{equation}
	\delta J_c^{\mu}=\frac{\gamma}{4}\sum_{x',\alpha,\alpha'}\Tr[\langle\gamma_{\alpha}^{\mu}\Psi_{\alpha}(x)M_{\alpha'}(x')\bar{\Psi}_{\alpha}(x)\rangle]\phi(x'),
\end{equation}
where $M_{\alpha'}(x'):=\partial(\bar{L}_{x'\alpha'}L_{x'\alpha'})/\partial\theta(x')|_{\theta=0}$ under the weak U(1) transformation~\eqref{eq: strong_U_1}. The nonvanishing $M_{\alpha'}(x)$ indicates that the density matrix does not remain block diagonal in the particle-number basis.  Therefore, the minimal condition for gauge invariance is the weak U(1) symmetry for a local Lindbladian under which $M_{\alpha'}(x)$ vanishes. A crucial question here is: how can we experimentally test our gauge-invariant response theory in the absence of the particle-number conservation?
%It is difficult to detect the breaking of weak U(1) symmetry since it only leads to the coherence in the density matrix, which cannot be measured in the gauge-invariant macroscopic observables.
Notice that the violation of the weak U(1) symmetry is associated with another Ward-Takahashi identity under the local weak U(1) transformation~\eqref{eq: strong_U_1}:
\begin{align}
	&[\delta (x - x_1) \tau_3C_{\alpha\alpha}(x_1,x_2) - \delta (x - x_2) C_{\alpha\alpha}(x_1,x_2)\tau_3] \nonumber\\
	&= i\langle \Psi_{\alpha} (x_1) \bar{\Psi}_{\alpha} (x_2)
	\partial_{\mu} \bar{J}_q^{\mu}(x) \rangle,
\end{align}
where the current $\bar{J}^{\mu}_q:=J^{\mu}_q+\Delta J^{\mu}_q$ is defined as
\begin{equation}\label{eq:Jq1}
	\begin{aligned}
		J_{q}^{\mu}&=J_{+}^{\mu}-J_{-}^{\mu},\\
		\partial_{\mu}\Delta J_q^{\mu}&=\!\frac{i\gamma}{2}\sum_{x'}\!\frac{\partial}{\partial\theta(x')}\!\!\!\left(2L_{x'+}\bar{L}_{x'-}-\sum_{\alpha}\bar{L}_{x'\alpha}L_{x'\alpha}\right)\!\!\Big|_{\theta=0}\!\!\!.
	\end{aligned}
\end{equation}
Here the non-vanishing current $\Delta J^{\mu}_q$ represents the fluctuation of currents between the contours and leads to the violation of gauge invariance when the weak U(1) symmetry is broken. The total current $\bar{J}^{\mu}_q$ is related to the conservation of the difference $n_{+}-n_{-}$ of the particle number between contours. However, the influence from the current $\Delta J_q^{\mu}$ in Eq. \eqref{eq:Jq1} cannot be detected directly by measuring $\langle N_{+}-N_{-}\rangle$ since $\langle N_{+}-N_{-}\rangle=0$ always holds. Instead, we propose to measure
\begin{align}\label{eq: OTOC}
	O_N(t)&:=\Tr[N\rho(t)N\rho(t)]-\Tr[N^2\rho(t)^2]\nonumber\\
	&=-\frac{1}{2}\langle\langle\rho|(N\otimes I-I\otimes N)^2|\rho\rangle\rangle,
\end{align}
where $|\rho\rangle\rangle$ is the vectorized density matrix $|\rho\ket\ket:=\sum_{i,j}\rho_{ij}|i\ket|j\ket$ for $\rho=\sum_{i,j}\rho_{ij}|i\ket\bra j|$. The operator $N\otimes I-I\otimes N$ corresponds to the difference $N_+-N_-$ in the path-integral representation. If the theory is gauge invariant, $O_N(t)$ is conserved since we can rewrite the observable as
\begin{equation}
	O_N(t)=\frac{1}{2}\Tr[[N,\rho(t)][N,\rho(t)]].
\end{equation}
When the Lindbladian has weak U(1) symmetry, we always have $[N,e^{\mathcal{L}t}\rho]=e^{\mathcal{L}t}[N,\rho]$. If the initial density matrix satisfies weak U(1) symmetry. i.e., $[N,\rho(0)]=0$, then $O_N(t)=0$ in the dynamics. Physically, a nonzero $O_N(t)$ indicates the superposition of states with different particle numbers and is related to quantum coherence of the system. Thus, the conservation of $O_N(t)$ serves as a necessary condition for the verification of gauge-invariant response theory. The main results are summarized in Table. \ref{table:relation}. 

We take the dissipative BCS superconductivity as an example to illustrate the equivalence between the conservation of $O_N$ and the gauge invariance~\cite{Yamamoto2021}. With mean-field approximation explicitly breaking the weak U(1) symmetry, one can show that $O_N(t)=-2N(t)+\sum_{\bm{k}}n_{\bm{k}}^2<0$ with $n_{\bm{k}}$ being the average particle number with momentum $\bm{k}$ (see Supplemental Material for detailed calculation~\cite{SupplementaryMaterial,mainfootnote3}). One can show that $O_N(t)$ is always negative and increases with time. However, due to the weak U(1) symmetry, the quantity $O_N(t)$ should be conserved.  Hence, the mean-field solution predicts unphysical non-conservation of $O_N$, indicating the necessity of our gauge-invariant theory to correctly calculate $O_N$, which describes quantum coherence of superconductors.  

\emph{Nambu-Goldstone mode.---}We consider low-energy collective modes of a weak U(1)-symmetric Lindbladian. For the case of three-dimensional dissipative BCS superconductivity without the EM field, the NG mode arises from spontaneous weak U(1) symmetry breaking. By applying the mean-field and quasi-steady-state approximations to the action \eqref{eq:Keldysh_action}, we obtain the recursive relation for the full vertex $\bar{\Gamma}^{\mu}$ as
%and obtain
%\begin{equation}
%%		B^{\dagger} & h_{-}^{2\times2}
%	\end{pmatrix}\Psi_{\tmmathbf{k},\omega},
%\end{equation}
%where $ \bar{\Psi}_{\bm{k},\omega}= (c_{k \uparrow +}, \bar{c}_{k \downarrow +}, c_{k
%	\uparrow -}, \bar{c}_{k \downarrow -})^T$ with $k:=(\bm{k},\omega)$, 
%\begin{equation}
%	h_{\pm}^{2\times2}=\begin{pmatrix}\pm(\omega-\varepsilon_{\bm{k}})+\frac{i\gamma n}{2} & -\Delta\\
%		-\Delta^{*} & \pm(\omega+\varepsilon_{\bm{k}})-\frac{i\gamma n}{2}
%	\end{pmatrix},
%\end{equation}
% $B=\begin{pmatrix}0 & 0\\
%	0 & i\gamma n
%	\end{pmatrix}$ with
%\begin{equation}
%	\Delta:=-(U_R+i\gamma/2)\int\frac{d^dk}{(2\pi)^d}\langle c_{-k\downarrow}c_{k\uparrow}\rangle
%\end{equation}
% being the order parameter of the dissipative system where $U_R$ is the strength of contact interaction. To see the NG mode, we rewrite the full vertex $\bar{\Gamma}$ as
\begin{equation} \label{eq: vertex}
	\bar{\Gamma}^{\mu}_{\alpha\delta}=\gamma_{\alpha \delta}^{\mu} + i \int dk \tau_3 (G (k + q) \bar{\Gamma}^{\mu} G (k))_{\delta\alpha} \tau_3 V_{\alpha\delta} (p -
	k) ,
\end{equation}
where $\int dk:=\int d^{d+1}k/(2\pi)^{d+1}$, $\gamma_{\alpha \delta}^{\mu}$ is the bare vertex and the interactions are given by
\begin{equation}
	V_{\pm \pm} = \pm U - i \gamma / 2, V_{- +} = i\gamma, V_{+-}=0,
\end{equation}
which are complex due to dissipation \footnote{In the Green's functions, we have taken the one-body loss channel into account to guarantee the iterative equation~\eqref{eq: vertex}.}. By considering the $\tau_2$ component of both sides of Eq. \eqref{eq: vertex}, we obtain the dispersion relation of the NG mode as
\begin{equation}
	\omega(\bm{k})=\pm v_s|\bm{k}|+iD|\bm{k}|^2,
\end{equation}
where $v_s=v_F/\sqrt{3}$ is the superfluid velocity for sound propagation with $v_F$ being the Fermi velocity and $D=3\sqrt{3}\gamma nv_F^2/(8\Delta^2)$ is the diffusion coefficient (see Supplemental Material~\cite{SupplementaryMaterial} for the derivation). We can see that the two-body loss induces diffusive propagation of the collective excitations, which is similar to the one in strong-to-weak U(1) symmetry breaking~\cite{Lessa2025,Minami2018,Xiaoyang2025} since we approximately have the strong U(1) symmetry under the quasi-steady-state approximation. Meanwhile, the linear-dispersion mode is attributed to the weak U(1) symmetry breaking as in closed quantum systems.
%\footnote{The analysis here is not restricted to fermionic systems and can be also applied to interacting Bose-Einstein condensates with two-body loss if we take the gauge invariance into account.}.

%Meanwhile, we also investigate the effective field theory of EM field. In the strong-interaction (BEC) limit, we can use the bosonic effective field theory to describe the many-body physics in this case. By integrating out the amplitude mode of the bosonic fields, we can see that the two-body loss also induces an effective electric field which refrains the particle loss~\cite{SupplementaryMaterial}.

\emph{Possible experimental situation.---} For ultracold atoms, the response current in Eq. \eqref{eq: response} for a dissipative superfluid can be measured by preparing a fermionic gas with particle loss in an optical lattice and applying an artificial electric field~\cite{Dahan1996,Wilkinson1996}. The electric current can thus be measured with single-bond resolution, which has already been realized in Refs.~ \cite{Impertro2024,Impertro2025}.  Meanwhile, the observable \eqref{eq: OTOC} can be measured in ultracold atomic systems by preparing two copies with the same initial states. Then the two parts in $O_N(t)$ can be measured by
\begin{align}\label{eq: ob1}
	\Tr[N\rho(t)N\rho(t)]=\sum_{\bm{x},\bm{y}}\Tr[(n(\bm{x})\otimes n(\bm{y}))\mathbb{S}(\rho\otimes\rho)],\\ \label{eq: ob2}
	\Tr[N^2\rho(t)^2]=\sum_{\bm{x},\bm{y}}\Tr[(n(\bm{x})n(\bm{y})\otimes I)\mathbb{S}(\rho\otimes\rho)],
\end{align}
where $\mathbb{S}$ is the SWAP operation~\cite{Klich2024} given by $\mathbb{S}=\sum_{ij}|i\rangle\langle j|\otimes|j\rangle\langle i|$ which transforms the state $|i\rangle|j\rangle$ into $|j\rangle|i\rangle$. By introducing weak tunneling between the two copies and Rabi oscillations~\cite{Haga2023}, one can measure the right-hand side of Eqs. \eqref{eq: ob1} and \eqref{eq: ob2} by Ramsey interferometry and finally measure $O_N(t)$. The observables \eqref{eq: ob1} and \eqref{eq: ob2} can also be measured by randomized measurement~\cite{Du2025,Kean2025,Elben2023,Notarnicola_2023}. 

\emph{Conclusion.---}In this Letter, we have developed the gauge-invariant response theory for dissipative fermionic systems in the presence of external EM gauge fields by establishing the Ward-Takahashi identity in open quantum systems and shown that the gauge invariance can be preserved even in the absence of particle-number conservation. We have demonstrated that the minimal condition for gauge invariance is weak U(1) symmetry for a local Lindbladian. We have constructed an observable that can be used to test the gauge-invariant transport theory, which can be detected from measuring the quantities on doubled copies. We have also shown the low-energy collective modes for dissipative BCS superconductivity, where two-body loss induces a diffusive mode for propagation. 

\emph{Acknowledgements.---} We are grateful to Xiaoqi Sun and Zongping Gong for fruitful discussion. H. L. is supported by Forefront Physics and Mathematics Program to Drive Transformation (FoPM), a World-leading Innovative Graduate Study (WINGS) Program, the University of Tokyo. H. L. also acknowledges JSPS KAKENHI (Grant No.~JP24KJ0824). M.N. is supported by JSPS KAKENHI Grant No.~JP24K16989. M.U. is supported by JSPS KAKENHI Grant No. JP22H01152 and the CREST program “Quantum Frontiers” of JST (Grand No. JPMJCR23I1).

\bibliographystyle{apsrev4-2}
\bibliography{MyNewCollection}

%apsrev4-2.bst 2019-01-14 (MD) hand-edited version of apsrev4-1.bst
%Control: key (0)
%Control: author (72) initials jnrlst
%Control: editor formatted (1) identically to author
%Control: production of article title (-1) disabled
%Control: page (0) single
%Control: year (1) truncated
%Control: production of eprint (0) enabled
\begin{thebibliography}{74}%
\makeatletter
\providecommand \@ifxundefined [1]{%
 \@ifx{#1\undefined}
}%
\providecommand \@ifnum [1]{%
 \ifnum #1\expandafter \@firstoftwo
 \else \expandafter \@secondoftwo
 \fi
}%
\providecommand \@ifx [1]{%
 \ifx #1\expandafter \@firstoftwo
 \else \expandafter \@secondoftwo
 \fi
}%
\providecommand \natexlab [1]{#1}%
\providecommand \enquote  [1]{``#1''}%
\providecommand \bibnamefont  [1]{#1}%
\providecommand \bibfnamefont [1]{#1}%
\providecommand \citenamefont [1]{#1}%
\providecommand \href@noop [0]{\@secondoftwo}%
\providecommand \href [0]{\begingroup \@sanitize@url \@href}%
\providecommand \@href[1]{\@@startlink{#1}\@@href}%
\providecommand \@@href[1]{\endgroup#1\@@endlink}%
\providecommand \@sanitize@url [0]{\catcode `\\12\catcode `\$12\catcode
  `\&12\catcode `\#12\catcode `\^12\catcode `\_12\catcode `\%12\relax}%
\providecommand \@@startlink[1]{}%
\providecommand \@@endlink[0]{}%
\providecommand \url  [0]{\begingroup\@sanitize@url \@url }%
\providecommand \@url [1]{\endgroup\@href {#1}{\urlprefix }}%
\providecommand \urlprefix  [0]{URL }%
\providecommand \Eprint [0]{\href }%
\providecommand \doibase [0]{https://doi.org/}%
\providecommand \selectlanguage [0]{\@gobble}%
\providecommand \bibinfo  [0]{\@secondoftwo}%
\providecommand \bibfield  [0]{\@secondoftwo}%
\providecommand \translation [1]{[#1]}%
\providecommand \BibitemOpen [0]{}%
\providecommand \bibitemStop [0]{}%
\providecommand \bibitemNoStop [0]{.\EOS\space}%
\providecommand \EOS [0]{\spacefactor3000\relax}%
\providecommand \BibitemShut  [1]{\csname bibitem#1\endcsname}%
\let\auto@bib@innerbib\@empty
%</preamble>
\bibitem [{\citenamefont {Leggett}(2006)}]{Leggett2006}%
  \BibitemOpen
  \bibfield  {author} {\bibinfo {author} {\bibfnamefont {A.~J.}\ \bibnamefont
  {Leggett}},\ }\href
  {https://doi.org/10.1093/acprof:oso/9780198526438.001.0001} {\emph {\bibinfo
  {title} {{Quantum Liquids: Bose condensation and Cooper pairing in
  condensed-matter systems}}}}\ (\bibinfo  {publisher} {Oxford University
  Press},\ \bibinfo {year} {2006})\BibitemShut {NoStop}%
\bibitem [{\citenamefont {Leggett}(1975)}]{RevModPhys.47.331}%
  \BibitemOpen
  \bibfield  {author} {\bibinfo {author} {\bibfnamefont {A.~J.}\ \bibnamefont
  {Leggett}},\ }\href {https://doi.org/10.1103/RevModPhys.47.331} {\bibfield
  {journal} {\bibinfo  {journal} {Rev. Mod. Phys.}\ }\textbf {\bibinfo {volume}
  {47}},\ \bibinfo {pages} {331} (\bibinfo {year} {1975})}\BibitemShut
  {NoStop}%
\bibitem [{\citenamefont {Coleman}(2015)}]{Coleman_2015}%
  \BibitemOpen
  \bibfield  {author} {\bibinfo {author} {\bibfnamefont {P.}~\bibnamefont
  {Coleman}},\ }\href@noop {} {\emph {\bibinfo {title} {Introduction to
  Many-Body Physics}}}\ (\bibinfo  {publisher} {Cambridge University Press},\
  \bibinfo {year} {2015})\BibitemShut {NoStop}%
\bibitem [{\citenamefont {Gross}(1996)}]{David_Symmetry}%
  \BibitemOpen
  \bibfield  {author} {\bibinfo {author} {\bibfnamefont {D.}~\bibnamefont
  {Gross}},\ }\href {https://doi.org/10.1073/pnas.93.25.14256} {\bibfield
  {journal} {\bibinfo  {journal} {Proceedings of the National Academy of
  Sciences}\ }\textbf {\bibinfo {volume} {93}},\ \bibinfo {pages} {14256}
  (\bibinfo {year} {1996})}\BibitemShut {NoStop}%
\bibitem [{\citenamefont {Ueda}(2010)}]{Ueda2010}%
  \BibitemOpen
  \bibfield  {author} {\bibinfo {author} {\bibfnamefont {M.}~\bibnamefont
  {Ueda}},\ }\href {https://doi.org/10.1142/7216} {\emph {\bibinfo {title}
  {Fundamentals and New Frontiers of Bose-Einstein Condensation}}}\ (\bibinfo
  {publisher} {WORLD SCIENTIFIC},\ \bibinfo {year} {2010})\BibitemShut
  {NoStop}%
\bibitem [{\citenamefont {Kibble}(2015)}]{Kibble_review}%
  \BibitemOpen
  \bibfield  {author} {\bibinfo {author} {\bibfnamefont {T.~W.~B.}\
  \bibnamefont {Kibble}},\ }\href {https://doi.org/10.1098/rsta.2014.0033}
  {\bibfield  {journal} {\bibinfo  {journal} {Philosophical Transactions of the
  Royal Society A: Mathematical, Physical and Engineering Sciences}\ }\textbf
  {\bibinfo {volume} {373}},\ \bibinfo {pages} {20140033} (\bibinfo {year}
  {2015})}\BibitemShut {NoStop}%
\bibitem [{\citenamefont {Lieu}\ \emph {et~al.}(2020)\citenamefont {Lieu},
  \citenamefont {Belyansky}, \citenamefont {Young}, \citenamefont {Lundgren},
  \citenamefont {Albert},\ and\ \citenamefont {Gorshkov}}]{Lieu2020}%
  \BibitemOpen
  \bibfield  {author} {\bibinfo {author} {\bibfnamefont {S.}~\bibnamefont
  {Lieu}}, \bibinfo {author} {\bibfnamefont {R.}~\bibnamefont {Belyansky}},
  \bibinfo {author} {\bibfnamefont {J.~T.}\ \bibnamefont {Young}}, \bibinfo
  {author} {\bibfnamefont {R.}~\bibnamefont {Lundgren}}, \bibinfo {author}
  {\bibfnamefont {V.~V.}\ \bibnamefont {Albert}},\ and\ \bibinfo {author}
  {\bibfnamefont {A.~V.}\ \bibnamefont {Gorshkov}},\ }\href
  {https://doi.org/10.1103/PhysRevLett.125.240405} {\bibfield  {journal}
  {\bibinfo  {journal} {Phys. Rev. Lett.}\ }\textbf {\bibinfo {volume} {125}},\
  \bibinfo {pages} {240405} (\bibinfo {year} {2020})}\BibitemShut {NoStop}%
\bibitem [{\citenamefont {Schmitt}(2015)}]{Schmitt2015}%
  \BibitemOpen
  \bibfield  {author} {\bibinfo {author} {\bibfnamefont {A.}~\bibnamefont
  {Schmitt}},\ }\href {https://doi.org/10.1007/978-3-319-07947-9} {\emph
  {\bibinfo {title} {Introduction to Superfluidity}}},\ Vol.\ \bibinfo {volume}
  {888}\ (\bibinfo  {publisher} {Springer International Publishing},\ \bibinfo
  {year} {2015})\BibitemShut {NoStop}%
\bibitem [{\citenamefont {Feynman}(1957)}]{RevModPhys.29.205}%
  \BibitemOpen
  \bibfield  {author} {\bibinfo {author} {\bibfnamefont {R.~P.}\ \bibnamefont
  {Feynman}},\ }\href {https://doi.org/10.1103/RevModPhys.29.205} {\bibfield
  {journal} {\bibinfo  {journal} {Rev. Mod. Phys.}\ }\textbf {\bibinfo {volume}
  {29}},\ \bibinfo {pages} {205} (\bibinfo {year} {1957})}\BibitemShut
  {NoStop}%
\bibitem [{\citenamefont {Bardeen}\ \emph {et~al.}(1957)\citenamefont
  {Bardeen}, \citenamefont {Cooper},\ and\ \citenamefont
  {Schrieffer}}]{BCS1957}%
  \BibitemOpen
  \bibfield  {author} {\bibinfo {author} {\bibfnamefont {J.}~\bibnamefont
  {Bardeen}}, \bibinfo {author} {\bibfnamefont {L.~N.}\ \bibnamefont
  {Cooper}},\ and\ \bibinfo {author} {\bibfnamefont {J.~R.}\ \bibnamefont
  {Schrieffer}},\ }\href {https://doi.org/10.1103/PhysRev.108.1175} {\bibfield
  {journal} {\bibinfo  {journal} {Phys. Rev.}\ }\textbf {\bibinfo {volume}
  {108}},\ \bibinfo {pages} {1175} (\bibinfo {year} {1957})}\BibitemShut
  {NoStop}%
\bibitem [{\citenamefont {Dalfovo}\ \emph {et~al.}(1999)\citenamefont
  {Dalfovo}, \citenamefont {Giorgini}, \citenamefont {Pitaevskii},\ and\
  \citenamefont {Stringari}}]{RevModPhys.71.463}%
  \BibitemOpen
  \bibfield  {author} {\bibinfo {author} {\bibfnamefont {F.}~\bibnamefont
  {Dalfovo}}, \bibinfo {author} {\bibfnamefont {S.}~\bibnamefont {Giorgini}},
  \bibinfo {author} {\bibfnamefont {L.~P.}\ \bibnamefont {Pitaevskii}},\ and\
  \bibinfo {author} {\bibfnamefont {S.}~\bibnamefont {Stringari}},\ }\href
  {https://doi.org/10.1103/RevModPhys.71.463} {\bibfield  {journal} {\bibinfo
  {journal} {Rev. Mod. Phys.}\ }\textbf {\bibinfo {volume} {71}},\ \bibinfo
  {pages} {463} (\bibinfo {year} {1999})}\BibitemShut {NoStop}%
\bibitem [{\citenamefont {Pethick}\ and\ \citenamefont
  {Smith}(2008)}]{Pethick_Smith_2008}%
  \BibitemOpen
  \bibfield  {author} {\bibinfo {author} {\bibfnamefont {C.~J.}\ \bibnamefont
  {Pethick}}\ and\ \bibinfo {author} {\bibfnamefont {H.}~\bibnamefont
  {Smith}},\ }\href@noop {} {\emph {\bibinfo {title} {Bose-Einstein
  Condensation in Dilute Gases}}},\ \bibinfo {edition} {2nd}\ ed.\ (\bibinfo
  {publisher} {Cambridge University Press},\ \bibinfo {year}
  {2008})\BibitemShut {NoStop}%
\bibitem [{\citenamefont {Anderson}(1958)}]{Anderson1958_SC}%
  \BibitemOpen
  \bibfield  {author} {\bibinfo {author} {\bibfnamefont {P.~W.}\ \bibnamefont
  {Anderson}},\ }\href {https://doi.org/10.1103/PhysRev.110.827} {\bibfield
  {journal} {\bibinfo  {journal} {Phys. Rev.}\ }\textbf {\bibinfo {volume}
  {110}},\ \bibinfo {pages} {827} (\bibinfo {year} {1958})}\BibitemShut
  {NoStop}%
\bibitem [{\citenamefont {London}(1948)}]{London1948}%
  \BibitemOpen
  \bibfield  {author} {\bibinfo {author} {\bibfnamefont {F.}~\bibnamefont
  {London}},\ }\href {https://doi.org/10.1103/PhysRev.74.562} {\bibfield
  {journal} {\bibinfo  {journal} {Phys. Rev.}\ }\textbf {\bibinfo {volume}
  {74}},\ \bibinfo {pages} {562} (\bibinfo {year} {1948})}\BibitemShut
  {NoStop}%
\bibitem [{\citenamefont {London}\ \emph {et~al.}(1935)\citenamefont {London},
  \citenamefont {London},\ and\ \citenamefont {Lindemann}}]{London1935}%
  \BibitemOpen
  \bibfield  {author} {\bibinfo {author} {\bibfnamefont {F.}~\bibnamefont
  {London}}, \bibinfo {author} {\bibfnamefont {H.}~\bibnamefont {London}},\
  and\ \bibinfo {author} {\bibfnamefont {F.~A.}\ \bibnamefont {Lindemann}},\
  }\href {https://doi.org/10.1098/rspa.1935.0048} {\bibfield  {journal}
  {\bibinfo  {journal} {Proceedings of the Royal Society of London. Series A -
  Mathematical and Physical Sciences}\ }\textbf {\bibinfo {volume} {149}},\
  \bibinfo {pages} {71} (\bibinfo {year} {1935})}\BibitemShut {NoStop}%
\bibitem [{\citenamefont {Bardeen}(1951)}]{Bardeen1951}%
  \BibitemOpen
  \bibfield  {author} {\bibinfo {author} {\bibfnamefont {J.}~\bibnamefont
  {Bardeen}},\ }\href {https://doi.org/10.1103/PhysRev.81.469.2} {\bibfield
  {journal} {\bibinfo  {journal} {Phys. Rev.}\ }\textbf {\bibinfo {volume}
  {81}},\ \bibinfo {pages} {469} (\bibinfo {year} {1951})}\BibitemShut
  {NoStop}%
\bibitem [{\citenamefont {Nambu}(1960)}]{Nambu1960}%
  \BibitemOpen
  \bibfield  {author} {\bibinfo {author} {\bibfnamefont {Y.}~\bibnamefont
  {Nambu}},\ }\href {https://doi.org/10.1103/PhysRev.117.648} {\bibfield
  {journal} {\bibinfo  {journal} {Phys. Rev.}\ }\textbf {\bibinfo {volume}
  {117}},\ \bibinfo {pages} {648} (\bibinfo {year} {1960})}\BibitemShut
  {NoStop}%
\bibitem [{\citenamefont {Ward}(1950)}]{Ward1950}%
  \BibitemOpen
  \bibfield  {author} {\bibinfo {author} {\bibfnamefont {J.~C.}\ \bibnamefont
  {Ward}},\ }\href {https://doi.org/10.1103/PhysRev.78.182} {\bibfield
  {journal} {\bibinfo  {journal} {Phys. Rev.}\ }\textbf {\bibinfo {volume}
  {78}},\ \bibinfo {pages} {182} (\bibinfo {year} {1950})}\BibitemShut
  {NoStop}%
\bibitem [{\citenamefont {Takahashi}(1957)}]{Takahashi1957}%
  \BibitemOpen
  \bibfield  {author} {\bibinfo {author} {\bibfnamefont {Y.}~\bibnamefont
  {Takahashi}},\ }\href {https://doi.org/10.1007/BF02832514} {\bibfield
  {journal} {\bibinfo  {journal} {Il Nuovo Cimento (1955-1965)}\ }\textbf
  {\bibinfo {volume} {6}},\ \bibinfo {pages} {371} (\bibinfo {year}
  {1957})}\BibitemShut {NoStop}%
\bibitem [{\citenamefont {Goldstone}(1961)}]{Goldstone_theorem}%
  \BibitemOpen
  \bibfield  {author} {\bibinfo {author} {\bibfnamefont {J.}~\bibnamefont
  {Goldstone}},\ }\href {https://doi.org/10.1007/BF02812722} {\bibfield
  {journal} {\bibinfo  {journal} {Il Nuovo Cimento (1955-1965)}\ }\textbf
  {\bibinfo {volume} {19}},\ \bibinfo {pages} {154} (\bibinfo {year}
  {1961})}\BibitemShut {NoStop}%
\bibitem [{\citenamefont {Goldstone}\ \emph {et~al.}(1962)\citenamefont
  {Goldstone}, \citenamefont {Salam},\ and\ \citenamefont
  {Weinberg}}]{Goldstone1962}%
  \BibitemOpen
  \bibfield  {author} {\bibinfo {author} {\bibfnamefont {J.}~\bibnamefont
  {Goldstone}}, \bibinfo {author} {\bibfnamefont {A.}~\bibnamefont {Salam}},\
  and\ \bibinfo {author} {\bibfnamefont {S.}~\bibnamefont {Weinberg}},\ }\href
  {https://doi.org/10.1103/PhysRev.127.965} {\bibfield  {journal} {\bibinfo
  {journal} {Phys. Rev.}\ }\textbf {\bibinfo {volume} {127}},\ \bibinfo {pages}
  {965} (\bibinfo {year} {1962})}\BibitemShut {NoStop}%
\bibitem [{\citenamefont {Ni}\ \emph {et~al.}(2008)\citenamefont {Ni},
  \citenamefont {Ospelkaus}, \citenamefont {de~Miranda}, \citenamefont {Pe'er},
  \citenamefont {Neyenhuis}, \citenamefont {Zirbel}, \citenamefont
  {Kotochigova}, \citenamefont {Julienne}, \citenamefont {Jin},\ and\
  \citenamefont {Ye}}]{Ni2008}%
  \BibitemOpen
  \bibfield  {author} {\bibinfo {author} {\bibfnamefont {K.-K.}\ \bibnamefont
  {Ni}}, \bibinfo {author} {\bibfnamefont {S.}~\bibnamefont {Ospelkaus}},
  \bibinfo {author} {\bibfnamefont {M.~H.~G.}\ \bibnamefont {de~Miranda}},
  \bibinfo {author} {\bibfnamefont {A.}~\bibnamefont {Pe'er}}, \bibinfo
  {author} {\bibfnamefont {B.}~\bibnamefont {Neyenhuis}}, \bibinfo {author}
  {\bibfnamefont {J.~J.}\ \bibnamefont {Zirbel}}, \bibinfo {author}
  {\bibfnamefont {S.}~\bibnamefont {Kotochigova}}, \bibinfo {author}
  {\bibfnamefont {P.~S.}\ \bibnamefont {Julienne}}, \bibinfo {author}
  {\bibfnamefont {D.~S.}\ \bibnamefont {Jin}},\ and\ \bibinfo {author}
  {\bibfnamefont {J.}~\bibnamefont {Ye}},\ }\href
  {https://doi.org/10.1126/science.1163861} {\bibfield  {journal} {\bibinfo
  {journal} {Science}\ }\textbf {\bibinfo {volume} {322}},\ \bibinfo {pages}
  {231} (\bibinfo {year} {2008})}\BibitemShut {NoStop}%
\bibitem [{\citenamefont {Ni}\ \emph {et~al.}(2010)\citenamefont {Ni},
  \citenamefont {Ospelkaus}, \citenamefont {Wang}, \citenamefont
  {Qu{\'e}m{\'e}ner}, \citenamefont {Neyenhuis}, \citenamefont {de~Miranda},
  \citenamefont {Bohn}, \citenamefont {Ye},\ and\ \citenamefont
  {Jin}}]{Ni2010}%
  \BibitemOpen
  \bibfield  {author} {\bibinfo {author} {\bibfnamefont {K.~K.}\ \bibnamefont
  {Ni}}, \bibinfo {author} {\bibfnamefont {S.}~\bibnamefont {Ospelkaus}},
  \bibinfo {author} {\bibfnamefont {D.}~\bibnamefont {Wang}}, \bibinfo {author}
  {\bibfnamefont {G.}~\bibnamefont {Qu{\'e}m{\'e}ner}}, \bibinfo {author}
  {\bibfnamefont {B.}~\bibnamefont {Neyenhuis}}, \bibinfo {author}
  {\bibfnamefont {M.~H.~G.}\ \bibnamefont {de~Miranda}}, \bibinfo {author}
  {\bibfnamefont {J.~L.}\ \bibnamefont {Bohn}}, \bibinfo {author}
  {\bibfnamefont {J.}~\bibnamefont {Ye}},\ and\ \bibinfo {author}
  {\bibfnamefont {D.~S.}\ \bibnamefont {Jin}},\ }\href
  {https://doi.org/10.1038/nature08953} {\bibfield  {journal} {\bibinfo
  {journal} {Nature}\ }\textbf {\bibinfo {volume} {464}},\ \bibinfo {pages}
  {1324} (\bibinfo {year} {2010})}\BibitemShut {NoStop}%
\bibitem [{\citenamefont {Ospelkaus}\ \emph {et~al.}(2010)\citenamefont
  {Ospelkaus}, \citenamefont {Ni}, \citenamefont {Wang}, \citenamefont
  {de~Miranda}, \citenamefont {Neyenhuis}, \citenamefont {Qu{\'e}m{\'e}ner},
  \citenamefont {Julienne}, \citenamefont {Bohn}, \citenamefont {Jin},\ and\
  \citenamefont {Ye}}]{Ospelkaus2010}%
  \BibitemOpen
  \bibfield  {author} {\bibinfo {author} {\bibfnamefont {S.}~\bibnamefont
  {Ospelkaus}}, \bibinfo {author} {\bibfnamefont {K.~K.}\ \bibnamefont {Ni}},
  \bibinfo {author} {\bibfnamefont {D.}~\bibnamefont {Wang}}, \bibinfo {author}
  {\bibfnamefont {M.~H.~G.}\ \bibnamefont {de~Miranda}}, \bibinfo {author}
  {\bibfnamefont {B.}~\bibnamefont {Neyenhuis}}, \bibinfo {author}
  {\bibfnamefont {G.}~\bibnamefont {Qu{\'e}m{\'e}ner}}, \bibinfo {author}
  {\bibfnamefont {P.~S.}\ \bibnamefont {Julienne}}, \bibinfo {author}
  {\bibfnamefont {J.~L.}\ \bibnamefont {Bohn}}, \bibinfo {author}
  {\bibfnamefont {D.~S.}\ \bibnamefont {Jin}},\ and\ \bibinfo {author}
  {\bibfnamefont {J.}~\bibnamefont {Ye}},\ }\href
  {https://doi.org/10.1126/science.1184121} {\bibfield  {journal} {\bibinfo
  {journal} {Science}\ }\textbf {\bibinfo {volume} {327}},\ \bibinfo {pages}
  {853} (\bibinfo {year} {2010})}\BibitemShut {NoStop}%
\bibitem [{\citenamefont {Liu}\ \emph {et~al.}(2020)\citenamefont {Liu},
  \citenamefont {Hu}, \citenamefont {Nichols}, \citenamefont {Grimes},
  \citenamefont {Karman}, \citenamefont {Guo},\ and\ \citenamefont
  {Ni}}]{Liu2020}%
  \BibitemOpen
  \bibfield  {author} {\bibinfo {author} {\bibfnamefont {Y.}~\bibnamefont
  {Liu}}, \bibinfo {author} {\bibfnamefont {M.-G.}\ \bibnamefont {Hu}},
  \bibinfo {author} {\bibfnamefont {M.~A.}\ \bibnamefont {Nichols}}, \bibinfo
  {author} {\bibfnamefont {D.~D.}\ \bibnamefont {Grimes}}, \bibinfo {author}
  {\bibfnamefont {T.}~\bibnamefont {Karman}}, \bibinfo {author} {\bibfnamefont
  {H.}~\bibnamefont {Guo}},\ and\ \bibinfo {author} {\bibfnamefont {K.-K.}\
  \bibnamefont {Ni}},\ }\href {https://doi.org/10.1038/s41567-020-0968-8}
  {\bibfield  {journal} {\bibinfo  {journal} {Nature Physics}\ }\textbf
  {\bibinfo {volume} {16}},\ \bibinfo {pages} {1132} (\bibinfo {year}
  {2020})}\BibitemShut {NoStop}%
\bibitem [{\citenamefont {Bause}\ \emph {et~al.}(2023)\citenamefont {Bause},
  \citenamefont {Christianen}, \citenamefont {Schindewolf}, \citenamefont
  {Bloch},\ and\ \citenamefont {Luo}}]{Bause2023}%
  \BibitemOpen
  \bibfield  {author} {\bibinfo {author} {\bibfnamefont {R.}~\bibnamefont
  {Bause}}, \bibinfo {author} {\bibfnamefont {A.}~\bibnamefont {Christianen}},
  \bibinfo {author} {\bibfnamefont {A.}~\bibnamefont {Schindewolf}}, \bibinfo
  {author} {\bibfnamefont {I.}~\bibnamefont {Bloch}},\ and\ \bibinfo {author}
  {\bibfnamefont {X.-Y.}\ \bibnamefont {Luo}},\ }\href
  {https://doi.org/10.1021/acs.jpca.2c08095} {\bibfield  {journal} {\bibinfo
  {journal} {The Journal of Physical Chemistry A}\ }\textbf {\bibinfo {volume}
  {127}},\ \bibinfo {pages} {729} (\bibinfo {year} {2023})}\BibitemShut
  {NoStop}%
\bibitem [{\citenamefont {Bause}\ \emph {et~al.}(2021)\citenamefont {Bause},
  \citenamefont {Schindewolf}, \citenamefont {Tao}, \citenamefont {Duda},
  \citenamefont {Chen}, \citenamefont {Qu\'em\'ener}, \citenamefont {Karman},
  \citenamefont {Christianen}, \citenamefont {Bloch},\ and\ \citenamefont
  {Luo}}]{Bause2021}%
  \BibitemOpen
  \bibfield  {author} {\bibinfo {author} {\bibfnamefont {R.}~\bibnamefont
  {Bause}}, \bibinfo {author} {\bibfnamefont {A.}~\bibnamefont {Schindewolf}},
  \bibinfo {author} {\bibfnamefont {R.}~\bibnamefont {Tao}}, \bibinfo {author}
  {\bibfnamefont {M.}~\bibnamefont {Duda}}, \bibinfo {author} {\bibfnamefont
  {X.-Y.}\ \bibnamefont {Chen}}, \bibinfo {author} {\bibfnamefont
  {G.}~\bibnamefont {Qu\'em\'ener}}, \bibinfo {author} {\bibfnamefont
  {T.}~\bibnamefont {Karman}}, \bibinfo {author} {\bibfnamefont
  {A.}~\bibnamefont {Christianen}}, \bibinfo {author} {\bibfnamefont
  {I.}~\bibnamefont {Bloch}},\ and\ \bibinfo {author} {\bibfnamefont {X.-Y.}\
  \bibnamefont {Luo}},\ }\href
  {https://doi.org/10.1103/PhysRevResearch.3.033013} {\bibfield  {journal}
  {\bibinfo  {journal} {Phys. Rev. Res.}\ }\textbf {\bibinfo {volume} {3}},\
  \bibinfo {pages} {033013} (\bibinfo {year} {2021})}\BibitemShut {NoStop}%
\bibitem [{\citenamefont {Will}\ \emph {et~al.}(2016)\citenamefont {Will},
  \citenamefont {Park}, \citenamefont {Yan}, \citenamefont {Loh},\ and\
  \citenamefont {Zwierlein}}]{Sebastian2016}%
  \BibitemOpen
  \bibfield  {author} {\bibinfo {author} {\bibfnamefont {S.~A.}\ \bibnamefont
  {Will}}, \bibinfo {author} {\bibfnamefont {J.~W.}\ \bibnamefont {Park}},
  \bibinfo {author} {\bibfnamefont {Z.~Z.}\ \bibnamefont {Yan}}, \bibinfo
  {author} {\bibfnamefont {H.}~\bibnamefont {Loh}},\ and\ \bibinfo {author}
  {\bibfnamefont {M.~W.}\ \bibnamefont {Zwierlein}},\ }\href
  {https://doi.org/10.1103/PhysRevLett.116.225306} {\bibfield  {journal}
  {\bibinfo  {journal} {Phys. Rev. Lett.}\ }\textbf {\bibinfo {volume} {116}},\
  \bibinfo {pages} {225306} (\bibinfo {year} {2016})}\BibitemShut {NoStop}%
\bibitem [{\citenamefont {Chen}\ \emph {et~al.}(2022)\citenamefont {Chen},
  \citenamefont {Duda}, \citenamefont {Schindewolf}, \citenamefont {Bause},
  \citenamefont {Bloch},\ and\ \citenamefont {Luo}}]{XingYan2022}%
  \BibitemOpen
  \bibfield  {author} {\bibinfo {author} {\bibfnamefont {X.-Y.}\ \bibnamefont
  {Chen}}, \bibinfo {author} {\bibfnamefont {M.}~\bibnamefont {Duda}}, \bibinfo
  {author} {\bibfnamefont {A.}~\bibnamefont {Schindewolf}}, \bibinfo {author}
  {\bibfnamefont {R.}~\bibnamefont {Bause}}, \bibinfo {author} {\bibfnamefont
  {I.}~\bibnamefont {Bloch}},\ and\ \bibinfo {author} {\bibfnamefont {X.-Y.}\
  \bibnamefont {Luo}},\ }\href {https://doi.org/10.1103/PhysRevLett.128.153401}
  {\bibfield  {journal} {\bibinfo  {journal} {Phys. Rev. Lett.}\ }\textbf
  {\bibinfo {volume} {128}},\ \bibinfo {pages} {153401} (\bibinfo {year}
  {2022})}\BibitemShut {NoStop}%
\bibitem [{\citenamefont {Yoshida}\ \emph {et~al.}(2018)\citenamefont
  {Yoshida}, \citenamefont {Saito}, \citenamefont {Waseem}, \citenamefont
  {Hattori},\ and\ \citenamefont {Mukaiyama}}]{Yoshida2018}%
  \BibitemOpen
  \bibfield  {author} {\bibinfo {author} {\bibfnamefont {J.}~\bibnamefont
  {Yoshida}}, \bibinfo {author} {\bibfnamefont {T.}~\bibnamefont {Saito}},
  \bibinfo {author} {\bibfnamefont {M.}~\bibnamefont {Waseem}}, \bibinfo
  {author} {\bibfnamefont {K.}~\bibnamefont {Hattori}},\ and\ \bibinfo {author}
  {\bibfnamefont {T.}~\bibnamefont {Mukaiyama}},\ }\href
  {https://doi.org/10.1103/PhysRevLett.120.133401} {\bibfield  {journal}
  {\bibinfo  {journal} {Phys. Rev. Lett.}\ }\textbf {\bibinfo {volume} {120}},\
  \bibinfo {pages} {133401} (\bibinfo {year} {2018})}\BibitemShut {NoStop}%
\bibitem [{\citenamefont {Helfrich}\ \emph {et~al.}(2010)\citenamefont
  {Helfrich}, \citenamefont {Hammer},\ and\ \citenamefont
  {Petrov}}]{Helfrich2010}%
  \BibitemOpen
  \bibfield  {author} {\bibinfo {author} {\bibfnamefont {K.}~\bibnamefont
  {Helfrich}}, \bibinfo {author} {\bibfnamefont {H.-W.}\ \bibnamefont
  {Hammer}},\ and\ \bibinfo {author} {\bibfnamefont {D.~S.}\ \bibnamefont
  {Petrov}},\ }\href {https://doi.org/10.1103/PhysRevA.81.042715} {\bibfield
  {journal} {\bibinfo  {journal} {Phys. Rev. A}\ }\textbf {\bibinfo {volume}
  {81}},\ \bibinfo {pages} {042715} (\bibinfo {year} {2010})}\BibitemShut
  {NoStop}%
\bibitem [{\citenamefont {Schindewolf}\ \emph {et~al.}(2022)\citenamefont
  {Schindewolf}, \citenamefont {Bause}, \citenamefont {Chen}, \citenamefont
  {Duda}, \citenamefont {Karman}, \citenamefont {Bloch},\ and\ \citenamefont
  {Luo}}]{Schindewolf2022}%
  \BibitemOpen
  \bibfield  {author} {\bibinfo {author} {\bibfnamefont {A.}~\bibnamefont
  {Schindewolf}}, \bibinfo {author} {\bibfnamefont {R.}~\bibnamefont {Bause}},
  \bibinfo {author} {\bibfnamefont {X.-Y.}\ \bibnamefont {Chen}}, \bibinfo
  {author} {\bibfnamefont {M.}~\bibnamefont {Duda}}, \bibinfo {author}
  {\bibfnamefont {T.}~\bibnamefont {Karman}}, \bibinfo {author} {\bibfnamefont
  {I.}~\bibnamefont {Bloch}},\ and\ \bibinfo {author} {\bibfnamefont {X.-Y.}\
  \bibnamefont {Luo}},\ }\href {https://doi.org/10.1038/s41586-022-04900-0}
  {\bibfield  {journal} {\bibinfo  {journal} {Nature}\ }\textbf {\bibinfo
  {volume} {607}},\ \bibinfo {pages} {677} (\bibinfo {year}
  {2022})}\BibitemShut {NoStop}%
\bibitem [{\citenamefont {Breuer}\ and\ \citenamefont
  {P}(2007)}]{Theory_Open2007}%
  \BibitemOpen
  \bibfield  {author} {\bibinfo {author} {\bibfnamefont {H.-P.}\ \bibnamefont
  {Breuer}}\ and\ \bibinfo {author} {\bibfnamefont {F.}~\bibnamefont {P}},\
  }\href {https://doi.org/10.1093/acprof:oso/9780199213900.001.0001} {\emph
  {\bibinfo {title} {\textit{The Theory of Open Quantum Systems}}}}\ (\bibinfo
  {publisher} {Oxford University Press},\ \bibinfo {year} {2007})\BibitemShut
  {NoStop}%
\bibitem [{\citenamefont {Kamenev}(2011)}]{Kamenev_2011}%
  \BibitemOpen
  \bibfield  {author} {\bibinfo {author} {\bibfnamefont {A.}~\bibnamefont
  {Kamenev}},\ }\href@noop {} {\emph {\bibinfo {title} {Field Theory of
  Non-Equilibrium Systems}}}\ (\bibinfo  {publisher} {Cambridge University
  Press},\ \bibinfo {year} {2011})\BibitemShut {NoStop}%
\bibitem [{\citenamefont {Aoki}\ \emph {et~al.}(2014)\citenamefont {Aoki},
  \citenamefont {Tsuji}, \citenamefont {Eckstein}, \citenamefont {Kollar},
  \citenamefont {Oka},\ and\ \citenamefont {Werner}}]{Aoki2014}%
  \BibitemOpen
  \bibfield  {author} {\bibinfo {author} {\bibfnamefont {H.}~\bibnamefont
  {Aoki}}, \bibinfo {author} {\bibfnamefont {N.}~\bibnamefont {Tsuji}},
  \bibinfo {author} {\bibfnamefont {M.}~\bibnamefont {Eckstein}}, \bibinfo
  {author} {\bibfnamefont {M.}~\bibnamefont {Kollar}}, \bibinfo {author}
  {\bibfnamefont {T.}~\bibnamefont {Oka}},\ and\ \bibinfo {author}
  {\bibfnamefont {P.}~\bibnamefont {Werner}},\ }\href
  {https://doi.org/10.1103/RevModPhys.86.779} {\bibfield  {journal} {\bibinfo
  {journal} {Rev. Mod. Phys.}\ }\textbf {\bibinfo {volume} {86}},\ \bibinfo
  {pages} {779} (\bibinfo {year} {2014})}\BibitemShut {NoStop}%
\bibitem [{\citenamefont {Yang}\ \emph {et~al.}(2021)\citenamefont {Yang},
  \citenamefont {Yang},\ and\ \citenamefont {Liu}}]{Qinghong2021}%
  \BibitemOpen
  \bibfield  {author} {\bibinfo {author} {\bibfnamefont {Q.}~\bibnamefont
  {Yang}}, \bibinfo {author} {\bibfnamefont {Z.}~\bibnamefont {Yang}},\ and\
  \bibinfo {author} {\bibfnamefont {D.~E.}\ \bibnamefont {Liu}},\ }\href
  {https://doi.org/10.1103/PhysRevB.104.014512} {\bibfield  {journal} {\bibinfo
   {journal} {Phys. Rev. B}\ }\textbf {\bibinfo {volume} {104}},\ \bibinfo
  {pages} {014512} (\bibinfo {year} {2021})}\BibitemShut {NoStop}%
\bibitem [{\citenamefont {Tsuji}\ \emph {et~al.}(2009)\citenamefont {Tsuji},
  \citenamefont {Oka},\ and\ \citenamefont {Aoki}}]{Tsuji2009}%
  \BibitemOpen
  \bibfield  {author} {\bibinfo {author} {\bibfnamefont {N.}~\bibnamefont
  {Tsuji}}, \bibinfo {author} {\bibfnamefont {T.}~\bibnamefont {Oka}},\ and\
  \bibinfo {author} {\bibfnamefont {H.}~\bibnamefont {Aoki}},\ }\href
  {https://doi.org/10.1103/PhysRevLett.103.047403} {\bibfield  {journal}
  {\bibinfo  {journal} {Phys. Rev. Lett.}\ }\textbf {\bibinfo {volume} {103}},\
  \bibinfo {pages} {047403} (\bibinfo {year} {2009})}\BibitemShut {NoStop}%
\bibitem [{\citenamefont {Li}\ \emph {et~al.}(2023)\citenamefont {Li},
  \citenamefont {Yu}, \citenamefont {Nakagawa},\ and\ \citenamefont
  {Ueda}}]{Hongchao2023}%
  \BibitemOpen
  \bibfield  {author} {\bibinfo {author} {\bibfnamefont {H.}~\bibnamefont
  {Li}}, \bibinfo {author} {\bibfnamefont {X.-H.}\ \bibnamefont {Yu}}, \bibinfo
  {author} {\bibfnamefont {M.}~\bibnamefont {Nakagawa}},\ and\ \bibinfo
  {author} {\bibfnamefont {M.}~\bibnamefont {Ueda}},\ }\href
  {https://doi.org/10.1103/PhysRevLett.131.216001} {\bibfield  {journal}
  {\bibinfo  {journal} {Phys. Rev. Lett.}\ }\textbf {\bibinfo {volume} {131}},\
  \bibinfo {pages} {216001} (\bibinfo {year} {2023})}\BibitemShut {NoStop}%
\bibitem [{\citenamefont {Yamamoto}\ \emph {et~al.}(2019)\citenamefont
  {Yamamoto}, \citenamefont {Nakagawa}, \citenamefont {Adachi}, \citenamefont
  {Takasan}, \citenamefont {Ueda},\ and\ \citenamefont
  {Kawakami}}]{Yamamoto2019}%
  \BibitemOpen
  \bibfield  {author} {\bibinfo {author} {\bibfnamefont {K.}~\bibnamefont
  {Yamamoto}}, \bibinfo {author} {\bibfnamefont {M.}~\bibnamefont {Nakagawa}},
  \bibinfo {author} {\bibfnamefont {K.}~\bibnamefont {Adachi}}, \bibinfo
  {author} {\bibfnamefont {K.}~\bibnamefont {Takasan}}, \bibinfo {author}
  {\bibfnamefont {M.}~\bibnamefont {Ueda}},\ and\ \bibinfo {author}
  {\bibfnamefont {N.}~\bibnamefont {Kawakami}},\ }\href
  {https://doi.org/10.1103/PhysRevLett.123.123601} {\bibfield  {journal}
  {\bibinfo  {journal} {Phys. Rev. Lett.}\ }\textbf {\bibinfo {volume} {123}},\
  \bibinfo {pages} {123601} (\bibinfo {year} {2019})}\BibitemShut {NoStop}%
\bibitem [{\citenamefont {Yamamoto}\ \emph {et~al.}(2021)\citenamefont
  {Yamamoto}, \citenamefont {Nakagawa}, \citenamefont {Tsuji}, \citenamefont
  {Ueda},\ and\ \citenamefont {Kawakami}}]{Yamamoto2021}%
  \BibitemOpen
  \bibfield  {author} {\bibinfo {author} {\bibfnamefont {K.}~\bibnamefont
  {Yamamoto}}, \bibinfo {author} {\bibfnamefont {M.}~\bibnamefont {Nakagawa}},
  \bibinfo {author} {\bibfnamefont {N.}~\bibnamefont {Tsuji}}, \bibinfo
  {author} {\bibfnamefont {M.}~\bibnamefont {Ueda}},\ and\ \bibinfo {author}
  {\bibfnamefont {N.}~\bibnamefont {Kawakami}},\ }\href
  {https://doi.org/10.1103/PhysRevLett.127.055301} {\bibfield  {journal}
  {\bibinfo  {journal} {Phys. Rev. Lett.}\ }\textbf {\bibinfo {volume} {127}},\
  \bibinfo {pages} {055301} (\bibinfo {year} {2021})}\BibitemShut {NoStop}%
\bibitem [{\citenamefont {Li}\ \emph {et~al.}(2025{\natexlab{a}})\citenamefont
  {Li}, \citenamefont {Yu}, \citenamefont {Nakagawa},\ and\ \citenamefont
  {Ueda}}]{Hongchao2024}%
  \BibitemOpen
  \bibfield  {author} {\bibinfo {author} {\bibfnamefont {H.}~\bibnamefont
  {Li}}, \bibinfo {author} {\bibfnamefont {X.-H.}\ \bibnamefont {Yu}}, \bibinfo
  {author} {\bibfnamefont {M.}~\bibnamefont {Nakagawa}},\ and\ \bibinfo
  {author} {\bibfnamefont {M.}~\bibnamefont {Ueda}},\ }\href
  {https://doi.org/10.1103/ww8r-sjhb} {\bibfield  {journal} {\bibinfo
  {journal} {Phys. Rev. Lett.}\ }\textbf {\bibinfo {volume} {135}},\ \bibinfo
  {pages} {166001} (\bibinfo {year} {2025}{\natexlab{a}})}\BibitemShut
  {NoStop}%
\bibitem [{\citenamefont {Han}\ \emph {et~al.}(2009)\citenamefont {Han},
  \citenamefont {Chan}, \citenamefont {Yi}, \citenamefont {Daley},
  \citenamefont {Diehl}, \citenamefont {Zoller},\ and\ \citenamefont
  {Duan}}]{Han2009}%
  \BibitemOpen
  \bibfield  {author} {\bibinfo {author} {\bibfnamefont {Y.-J.}\ \bibnamefont
  {Han}}, \bibinfo {author} {\bibfnamefont {Y.-H.}\ \bibnamefont {Chan}},
  \bibinfo {author} {\bibfnamefont {W.}~\bibnamefont {Yi}}, \bibinfo {author}
  {\bibfnamefont {A.~J.}\ \bibnamefont {Daley}}, \bibinfo {author}
  {\bibfnamefont {S.}~\bibnamefont {Diehl}}, \bibinfo {author} {\bibfnamefont
  {P.}~\bibnamefont {Zoller}},\ and\ \bibinfo {author} {\bibfnamefont {L.-M.}\
  \bibnamefont {Duan}},\ }\href
  {https://doi.org/10.1103/PhysRevLett.103.070404} {\bibfield  {journal}
  {\bibinfo  {journal} {Phys. Rev. Lett.}\ }\textbf {\bibinfo {volume} {103}},\
  \bibinfo {pages} {070404} (\bibinfo {year} {2009})}\BibitemShut {NoStop}%
\bibitem [{\citenamefont {Diehl}\ \emph {et~al.}(2010)\citenamefont {Diehl},
  \citenamefont {Tomadin}, \citenamefont {Micheli}, \citenamefont {Fazio},\
  and\ \citenamefont {Zoller}}]{Diehl2010}%
  \BibitemOpen
  \bibfield  {author} {\bibinfo {author} {\bibfnamefont {S.}~\bibnamefont
  {Diehl}}, \bibinfo {author} {\bibfnamefont {A.}~\bibnamefont {Tomadin}},
  \bibinfo {author} {\bibfnamefont {A.}~\bibnamefont {Micheli}}, \bibinfo
  {author} {\bibfnamefont {R.}~\bibnamefont {Fazio}},\ and\ \bibinfo {author}
  {\bibfnamefont {P.}~\bibnamefont {Zoller}},\ }\href
  {https://doi.org/10.1103/PhysRevLett.105.015702} {\bibfield  {journal}
  {\bibinfo  {journal} {Phys. Rev. Lett.}\ }\textbf {\bibinfo {volume} {105}},\
  \bibinfo {pages} {015702} (\bibinfo {year} {2010})}\BibitemShut {NoStop}%
\bibitem [{\citenamefont {Sieberer}\ \emph {et~al.}(2016)\citenamefont
  {Sieberer}, \citenamefont {Buchhold},\ and\ \citenamefont
  {Diehl}}]{Sieberer_2016}%
  \BibitemOpen
  \bibfield  {author} {\bibinfo {author} {\bibfnamefont {L.~M.}\ \bibnamefont
  {Sieberer}}, \bibinfo {author} {\bibfnamefont {M.}~\bibnamefont {Buchhold}},\
  and\ \bibinfo {author} {\bibfnamefont {S.}~\bibnamefont {Diehl}},\ }\href
  {https://doi.org/10.1088/0034-4885/79/9/096001} {\bibfield  {journal}
  {\bibinfo  {journal} {Reports on Progress in Physics}\ }\textbf {\bibinfo
  {volume} {79}},\ \bibinfo {pages} {096001} (\bibinfo {year}
  {2016})}\BibitemShut {NoStop}%
\bibitem [{\citenamefont {Nakagawa}\ \emph {et~al.}(2021)\citenamefont
  {Nakagawa}, \citenamefont {Kawakami},\ and\ \citenamefont
  {Ueda}}]{Nakagawa2021}%
  \BibitemOpen
  \bibfield  {author} {\bibinfo {author} {\bibfnamefont {M.}~\bibnamefont
  {Nakagawa}}, \bibinfo {author} {\bibfnamefont {N.}~\bibnamefont {Kawakami}},\
  and\ \bibinfo {author} {\bibfnamefont {M.}~\bibnamefont {Ueda}},\ }\href
  {https://doi.org/10.1103/PhysRevLett.126.110404} {\bibfield  {journal}
  {\bibinfo  {journal} {Phys. Rev. Lett.}\ }\textbf {\bibinfo {volume} {126}},\
  \bibinfo {pages} {110404} (\bibinfo {year} {2021})}\BibitemShut {NoStop}%
\bibitem [{\citenamefont {Yamamoto}\ \emph {et~al.}(2020)\citenamefont
  {Yamamoto}, \citenamefont {Ashida},\ and\ \citenamefont
  {Kawakami}}]{Yamamoto2020Re}%
  \BibitemOpen
  \bibfield  {author} {\bibinfo {author} {\bibfnamefont {K.}~\bibnamefont
  {Yamamoto}}, \bibinfo {author} {\bibfnamefont {Y.}~\bibnamefont {Ashida}},\
  and\ \bibinfo {author} {\bibfnamefont {N.}~\bibnamefont {Kawakami}},\ }\href
  {https://doi.org/10.1103/PhysRevResearch.2.043343} {\bibfield  {journal}
  {\bibinfo  {journal} {Phys. Rev. Res.}\ }\textbf {\bibinfo {volume} {2}},\
  \bibinfo {pages} {043343} (\bibinfo {year} {2020})}\BibitemShut {NoStop}%
\bibitem [{\citenamefont {Damanet}\ \emph {et~al.}(2019)\citenamefont
  {Damanet}, \citenamefont {Mascarenhas}, \citenamefont {Pekker},\ and\
  \citenamefont {Daley}}]{Damanet2019}%
  \BibitemOpen
  \bibfield  {author} {\bibinfo {author} {\bibfnamefont {F.}~\bibnamefont
  {Damanet}}, \bibinfo {author} {\bibfnamefont {E.}~\bibnamefont
  {Mascarenhas}}, \bibinfo {author} {\bibfnamefont {D.}~\bibnamefont
  {Pekker}},\ and\ \bibinfo {author} {\bibfnamefont {A.~J.}\ \bibnamefont
  {Daley}},\ }\href {https://doi.org/10.1103/PhysRevLett.123.180402} {\bibfield
   {journal} {\bibinfo  {journal} {Phys. Rev. Lett.}\ }\textbf {\bibinfo
  {volume} {123}},\ \bibinfo {pages} {180402} (\bibinfo {year}
  {2019})}\BibitemShut {NoStop}%
\bibitem [{\citenamefont {Li}\ \emph {et~al.}(2025{\natexlab{b}})\citenamefont
  {Li}, \citenamefont {Shang}, \citenamefont {Kuwahara},\ and\ \citenamefont
  {Vu}}]{Hongchao2025}%
  \BibitemOpen
  \bibfield  {author} {\bibinfo {author} {\bibfnamefont {H.}~\bibnamefont
  {Li}}, \bibinfo {author} {\bibfnamefont {C.}~\bibnamefont {Shang}}, \bibinfo
  {author} {\bibfnamefont {T.}~\bibnamefont {Kuwahara}},\ and\ \bibinfo
  {author} {\bibfnamefont {T.~V.}\ \bibnamefont {Vu}},\ }\href
  {https://arxiv.org/abs/2503.13731} {\bibinfo {title} {Macroscopic particle
  transport in dissipative long-range bosonic systems}} (\bibinfo {year}
  {2025}{\natexlab{b}}),\ \Eprint {https://arxiv.org/abs/2503.13731}
  {arXiv:2503.13731 [quant-ph]} \BibitemShut {NoStop}%
\bibitem [{\citenamefont {Albert}\ and\ \citenamefont
  {Jiang}(2014)}]{Albert2014}%
  \BibitemOpen
  \bibfield  {author} {\bibinfo {author} {\bibfnamefont {V.~V.}\ \bibnamefont
  {Albert}}\ and\ \bibinfo {author} {\bibfnamefont {L.}~\bibnamefont {Jiang}},\
  }\href {https://doi.org/10.1103/PhysRevA.89.022118} {\bibfield  {journal}
  {\bibinfo  {journal} {Phys. Rev. A}\ }\textbf {\bibinfo {volume} {89}},\
  \bibinfo {pages} {022118} (\bibinfo {year} {2014})}\BibitemShut {NoStop}%
\bibitem [{\citenamefont {Daley}\ \emph {et~al.}(2012)\citenamefont {Daley},
  \citenamefont {Pichler}, \citenamefont {Schachenmayer},\ and\ \citenamefont
  {Zoller}}]{Daley2012}%
  \BibitemOpen
  \bibfield  {author} {\bibinfo {author} {\bibfnamefont {A.~J.}\ \bibnamefont
  {Daley}}, \bibinfo {author} {\bibfnamefont {H.}~\bibnamefont {Pichler}},
  \bibinfo {author} {\bibfnamefont {J.}~\bibnamefont {Schachenmayer}},\ and\
  \bibinfo {author} {\bibfnamefont {P.}~\bibnamefont {Zoller}},\ }\href
  {https://doi.org/10.1103/PhysRevLett.109.020505} {\bibfield  {journal}
  {\bibinfo  {journal} {Phys. Rev. Lett.}\ }\textbf {\bibinfo {volume} {109}},\
  \bibinfo {pages} {020505} (\bibinfo {year} {2012})}\BibitemShut {NoStop}%
\bibitem [{\citenamefont {Islam}\ \emph {et~al.}(2015)\citenamefont {Islam},
  \citenamefont {Ma}, \citenamefont {Preiss}, \citenamefont {Eric~Tai},
  \citenamefont {Lukin}, \citenamefont {Rispoli},\ and\ \citenamefont
  {Greiner}}]{Islam2015}%
  \BibitemOpen
  \bibfield  {author} {\bibinfo {author} {\bibfnamefont {R.}~\bibnamefont
  {Islam}}, \bibinfo {author} {\bibfnamefont {R.}~\bibnamefont {Ma}}, \bibinfo
  {author} {\bibfnamefont {P.~M.}\ \bibnamefont {Preiss}}, \bibinfo {author}
  {\bibfnamefont {M.}~\bibnamefont {Eric~Tai}}, \bibinfo {author}
  {\bibfnamefont {A.}~\bibnamefont {Lukin}}, \bibinfo {author} {\bibfnamefont
  {M.}~\bibnamefont {Rispoli}},\ and\ \bibinfo {author} {\bibfnamefont
  {M.}~\bibnamefont {Greiner}},\ }\href {https://doi.org/10.1038/nature15750}
  {\bibfield  {journal} {\bibinfo  {journal} {Nature}\ }\textbf {\bibinfo
  {volume} {528}},\ \bibinfo {pages} {77} (\bibinfo {year} {2015})}\BibitemShut
  {NoStop}%
\bibitem [{\citenamefont {Pichler}\ \emph {et~al.}(2016)\citenamefont
  {Pichler}, \citenamefont {Zhu}, \citenamefont {Seif}, \citenamefont
  {Zoller},\ and\ \citenamefont {Hafezi}}]{Pichler2016}%
  \BibitemOpen
  \bibfield  {author} {\bibinfo {author} {\bibfnamefont {H.}~\bibnamefont
  {Pichler}}, \bibinfo {author} {\bibfnamefont {G.}~\bibnamefont {Zhu}},
  \bibinfo {author} {\bibfnamefont {A.}~\bibnamefont {Seif}}, \bibinfo {author}
  {\bibfnamefont {P.}~\bibnamefont {Zoller}},\ and\ \bibinfo {author}
  {\bibfnamefont {M.}~\bibnamefont {Hafezi}},\ }\href
  {https://doi.org/10.1103/PhysRevX.6.041033} {\bibfield  {journal} {\bibinfo
  {journal} {Phys. Rev. X}\ }\textbf {\bibinfo {volume} {6}},\ \bibinfo {pages}
  {041033} (\bibinfo {year} {2016})}\BibitemShut {NoStop}%
\bibitem [{\citenamefont {Minami}\ and\ \citenamefont
  {Hidaka}(2018)}]{Minami2018}%
  \BibitemOpen
  \bibfield  {author} {\bibinfo {author} {\bibfnamefont {Y.}~\bibnamefont
  {Minami}}\ and\ \bibinfo {author} {\bibfnamefont {Y.}~\bibnamefont
  {Hidaka}},\ }\href {https://doi.org/10.1103/PhysRevE.97.012130} {\bibfield
  {journal} {\bibinfo  {journal} {Phys. Rev. E}\ }\textbf {\bibinfo {volume}
  {97}},\ \bibinfo {pages} {012130} (\bibinfo {year} {2018})}\BibitemShut
  {NoStop}%
\bibitem [{\citenamefont {Hidaka}\ and\ \citenamefont
  {Minami}(2020)}]{Hidaka2020}%
  \BibitemOpen
  \bibfield  {author} {\bibinfo {author} {\bibfnamefont {Y.}~\bibnamefont
  {Hidaka}}\ and\ \bibinfo {author} {\bibfnamefont {Y.}~\bibnamefont
  {Minami}},\ }\href {https://doi.org/10.1093/ptep/ptaa005} {\bibfield
  {journal} {\bibinfo  {journal} {Progress of Theoretical and Experimental
  Physics}\ }\textbf {\bibinfo {volume} {2020}},\ \bibinfo {pages} {033A01}
  (\bibinfo {year} {2020})}\BibitemShut {NoStop}%
\bibitem [{\citenamefont {Buča}\ and\ \citenamefont
  {Prosen}(2012)}]{Buca_2012}%
  \BibitemOpen
  \bibfield  {author} {\bibinfo {author} {\bibfnamefont {B.}~\bibnamefont
  {Buča}}\ and\ \bibinfo {author} {\bibfnamefont {T.}~\bibnamefont {Prosen}},\
  }\href {https://doi.org/10.1088/1367-2630/14/7/073007} {\bibfield  {journal}
  {\bibinfo  {journal} {New Journal of Physics}\ }\textbf {\bibinfo {volume}
  {14}},\ \bibinfo {pages} {073007} (\bibinfo {year} {2012})}\BibitemShut
  {NoStop}%
\bibitem [{\citenamefont {Gebauer}\ and\ \citenamefont
  {Car}(2004)}]{PhysRevLett.93.160404}%
  \BibitemOpen
  \bibfield  {author} {\bibinfo {author} {\bibfnamefont {R.}~\bibnamefont
  {Gebauer}}\ and\ \bibinfo {author} {\bibfnamefont {R.}~\bibnamefont {Car}},\
  }\href {https://doi.org/10.1103/PhysRevLett.93.160404} {\bibfield  {journal}
  {\bibinfo  {journal} {Phys. Rev. Lett.}\ }\textbf {\bibinfo {volume} {93}},\
  \bibinfo {pages} {160404} (\bibinfo {year} {2004})}\BibitemShut {NoStop}%
\bibitem [{Sup()}]{SupplementaryMaterial}%
  \BibitemOpen
  \href@noop {} {\bibinfo {title} {See supplemental material for
  details.}}\BibitemShut {Stop}%
\bibitem [{\citenamefont {Mazza}\ and\ \citenamefont
  {Schir\`o}(2023)}]{Mazza2023}%
  \BibitemOpen
  \bibfield  {author} {\bibinfo {author} {\bibfnamefont {G.}~\bibnamefont
  {Mazza}}\ and\ \bibinfo {author} {\bibfnamefont {M.}~\bibnamefont
  {Schir\`o}},\ }\href {https://doi.org/10.1103/PhysRevA.107.L051301}
  {\bibfield  {journal} {\bibinfo  {journal} {Phys. Rev. A}\ }\textbf {\bibinfo
  {volume} {107}},\ \bibinfo {pages} {L051301} (\bibinfo {year}
  {2023})}\BibitemShut {NoStop}%
\bibitem [{mai()}]{mainfootnote3}%
  \BibitemOpen
  \href@noop {} {\bibinfo {title} {We note that the one-body loss channel
  introduced in ref. \cite{Mazza2023} does not qualitatively change this
  result.}}\BibitemShut {Stop}%
\bibitem [{Note1()}]{Note1}%
  \BibitemOpen
  \bibinfo {note} {In the Green's functions, we have taken the one-body loss
  channel into account to guarantee the iterative equation~\protect \textup
  {\hbox {\mathsurround \z@ \protect \normalfont (\ignorespaces \ref {eq:
  vertex}\unskip \@@italiccorr )}}.}\BibitemShut {Stop}%
\bibitem [{\citenamefont {Lessa}\ \emph {et~al.}(2025)\citenamefont {Lessa},
  \citenamefont {Ma}, \citenamefont {Zhang}, \citenamefont {Bi}, \citenamefont
  {Cheng},\ and\ \citenamefont {Wang}}]{Lessa2025}%
  \BibitemOpen
  \bibfield  {author} {\bibinfo {author} {\bibfnamefont {L.~A.}\ \bibnamefont
  {Lessa}}, \bibinfo {author} {\bibfnamefont {R.}~\bibnamefont {Ma}}, \bibinfo
  {author} {\bibfnamefont {J.-H.}\ \bibnamefont {Zhang}}, \bibinfo {author}
  {\bibfnamefont {Z.}~\bibnamefont {Bi}}, \bibinfo {author} {\bibfnamefont
  {M.}~\bibnamefont {Cheng}},\ and\ \bibinfo {author} {\bibfnamefont
  {C.}~\bibnamefont {Wang}},\ }\href
  {https://doi.org/10.1103/PRXQuantum.6.010344} {\bibfield  {journal} {\bibinfo
   {journal} {PRX Quantum}\ }\textbf {\bibinfo {volume} {6}},\ \bibinfo {pages}
  {010344} (\bibinfo {year} {2025})}\BibitemShut {NoStop}%
\bibitem [{\citenamefont {Huang}\ \emph {et~al.}(2025)\citenamefont {Huang},
  \citenamefont {Qi}, \citenamefont {Zhang},\ and\ \citenamefont
  {Lucas}}]{Xiaoyang2025}%
  \BibitemOpen
  \bibfield  {author} {\bibinfo {author} {\bibfnamefont {X.}~\bibnamefont
  {Huang}}, \bibinfo {author} {\bibfnamefont {M.}~\bibnamefont {Qi}}, \bibinfo
  {author} {\bibfnamefont {J.-H.}\ \bibnamefont {Zhang}},\ and\ \bibinfo
  {author} {\bibfnamefont {A.}~\bibnamefont {Lucas}},\ }\href
  {https://doi.org/10.1103/PhysRevB.111.125147} {\bibfield  {journal} {\bibinfo
   {journal} {Phys. Rev. B}\ }\textbf {\bibinfo {volume} {111}},\ \bibinfo
  {pages} {125147} (\bibinfo {year} {2025})}\BibitemShut {NoStop}%
\bibitem [{\citenamefont {Ben~Dahan}\ \emph {et~al.}(1996)\citenamefont
  {Ben~Dahan}, \citenamefont {Peik}, \citenamefont {Reichel}, \citenamefont
  {Castin},\ and\ \citenamefont {Salomon}}]{Dahan1996}%
  \BibitemOpen
  \bibfield  {author} {\bibinfo {author} {\bibfnamefont {M.}~\bibnamefont
  {Ben~Dahan}}, \bibinfo {author} {\bibfnamefont {E.}~\bibnamefont {Peik}},
  \bibinfo {author} {\bibfnamefont {J.}~\bibnamefont {Reichel}}, \bibinfo
  {author} {\bibfnamefont {Y.}~\bibnamefont {Castin}},\ and\ \bibinfo {author}
  {\bibfnamefont {C.}~\bibnamefont {Salomon}},\ }\href
  {https://doi.org/10.1103/PhysRevLett.76.4508} {\bibfield  {journal} {\bibinfo
   {journal} {Phys. Rev. Lett.}\ }\textbf {\bibinfo {volume} {76}},\ \bibinfo
  {pages} {4508} (\bibinfo {year} {1996})}\BibitemShut {NoStop}%
\bibitem [{\citenamefont {Wilkinson}\ \emph {et~al.}(1996)\citenamefont
  {Wilkinson}, \citenamefont {Bharucha}, \citenamefont {Madison}, \citenamefont
  {Niu},\ and\ \citenamefont {Raizen}}]{Wilkinson1996}%
  \BibitemOpen
  \bibfield  {author} {\bibinfo {author} {\bibfnamefont {S.~R.}\ \bibnamefont
  {Wilkinson}}, \bibinfo {author} {\bibfnamefont {C.~F.}\ \bibnamefont
  {Bharucha}}, \bibinfo {author} {\bibfnamefont {K.~W.}\ \bibnamefont
  {Madison}}, \bibinfo {author} {\bibfnamefont {Q.}~\bibnamefont {Niu}},\ and\
  \bibinfo {author} {\bibfnamefont {M.~G.}\ \bibnamefont {Raizen}},\ }\href
  {https://doi.org/10.1103/PhysRevLett.76.4512} {\bibfield  {journal} {\bibinfo
   {journal} {Phys. Rev. Lett.}\ }\textbf {\bibinfo {volume} {76}},\ \bibinfo
  {pages} {4512} (\bibinfo {year} {1996})}\BibitemShut {NoStop}%
\bibitem [{\citenamefont {Impertro}\ \emph {et~al.}(2024)\citenamefont
  {Impertro}, \citenamefont {Karch}, \citenamefont {Wienand}, \citenamefont
  {Huh}, \citenamefont {Schweizer}, \citenamefont {Bloch},\ and\ \citenamefont
  {Aidelsburger}}]{Impertro2024}%
  \BibitemOpen
  \bibfield  {author} {\bibinfo {author} {\bibfnamefont {A.}~\bibnamefont
  {Impertro}}, \bibinfo {author} {\bibfnamefont {S.}~\bibnamefont {Karch}},
  \bibinfo {author} {\bibfnamefont {J.~F.}\ \bibnamefont {Wienand}}, \bibinfo
  {author} {\bibfnamefont {S.}~\bibnamefont {Huh}}, \bibinfo {author}
  {\bibfnamefont {C.}~\bibnamefont {Schweizer}}, \bibinfo {author}
  {\bibfnamefont {I.}~\bibnamefont {Bloch}},\ and\ \bibinfo {author}
  {\bibfnamefont {M.}~\bibnamefont {Aidelsburger}},\ }\href
  {https://doi.org/10.1103/PhysRevLett.133.063401} {\bibfield  {journal}
  {\bibinfo  {journal} {Phys. Rev. Lett.}\ }\textbf {\bibinfo {volume} {133}},\
  \bibinfo {pages} {063401} (\bibinfo {year} {2024})}\BibitemShut {NoStop}%
\bibitem [{\citenamefont {Impertro}\ \emph {et~al.}(2025)\citenamefont
  {Impertro}, \citenamefont {Huh}, \citenamefont {Karch}, \citenamefont
  {Wienand}, \citenamefont {Bloch},\ and\ \citenamefont
  {Aidelsburger}}]{Impertro2025}%
  \BibitemOpen
  \bibfield  {author} {\bibinfo {author} {\bibfnamefont {A.}~\bibnamefont
  {Impertro}}, \bibinfo {author} {\bibfnamefont {S.}~\bibnamefont {Huh}},
  \bibinfo {author} {\bibfnamefont {S.}~\bibnamefont {Karch}}, \bibinfo
  {author} {\bibfnamefont {J.~F.}\ \bibnamefont {Wienand}}, \bibinfo {author}
  {\bibfnamefont {I.}~\bibnamefont {Bloch}},\ and\ \bibinfo {author}
  {\bibfnamefont {M.}~\bibnamefont {Aidelsburger}},\ }\href
  {https://doi.org/10.1038/s41567-025-02890-0} {\bibfield  {journal} {\bibinfo
  {journal} {Nature Physics}\ }\textbf {\bibinfo {volume} {21}},\ \bibinfo
  {pages} {895} (\bibinfo {year} {2025})}\BibitemShut {NoStop}%
\bibitem [{\citenamefont {Klich}(2024)}]{Klich2024}%
  \BibitemOpen
  \bibfield  {author} {\bibinfo {author} {\bibfnamefont {I.}~\bibnamefont
  {Klich}},\ }\href {https://arxiv.org/abs/2408.15161} {\bibinfo {title} {Swap
  and transpose by displacements, stabilizer renyi entropies for continuous
  variables and qudits and other applications}} (\bibinfo {year} {2024}),\
  \Eprint {https://arxiv.org/abs/2408.15161} {arXiv:2408.15161 [quant-ph]}
  \BibitemShut {NoStop}%
\bibitem [{\citenamefont {Haga}\ \emph {et~al.}(2023)\citenamefont {Haga},
  \citenamefont {Nakagawa}, \citenamefont {Hamazaki},\ and\ \citenamefont
  {Ueda}}]{Haga2023}%
  \BibitemOpen
  \bibfield  {author} {\bibinfo {author} {\bibfnamefont {T.}~\bibnamefont
  {Haga}}, \bibinfo {author} {\bibfnamefont {M.}~\bibnamefont {Nakagawa}},
  \bibinfo {author} {\bibfnamefont {R.}~\bibnamefont {Hamazaki}},\ and\
  \bibinfo {author} {\bibfnamefont {M.}~\bibnamefont {Ueda}},\ }\href
  {https://doi.org/10.1103/PhysRevResearch.5.043225} {\bibfield  {journal}
  {\bibinfo  {journal} {Phys. Rev. Res.}\ }\textbf {\bibinfo {volume} {5}},\
  \bibinfo {pages} {043225} (\bibinfo {year} {2023})}\BibitemShut {NoStop}%
\bibitem [{\citenamefont {Du}\ \emph {et~al.}(2025)\citenamefont {Du},
  \citenamefont {Tang}, \citenamefont {Elben}, \citenamefont {Roth},
  \citenamefont {Eisert},\ and\ \citenamefont {Liu}}]{Du2025}%
  \BibitemOpen
  \bibfield  {author} {\bibinfo {author} {\bibfnamefont {Z.}~\bibnamefont
  {Du}}, \bibinfo {author} {\bibfnamefont {Y.}~\bibnamefont {Tang}}, \bibinfo
  {author} {\bibfnamefont {A.}~\bibnamefont {Elben}}, \bibinfo {author}
  {\bibfnamefont {I.}~\bibnamefont {Roth}}, \bibinfo {author} {\bibfnamefont
  {J.}~\bibnamefont {Eisert}},\ and\ \bibinfo {author} {\bibfnamefont
  {Z.}~\bibnamefont {Liu}},\ }\href {https://arxiv.org/abs/2505.09206}
  {\bibinfo {title} {Optimal randomized measurements for a family of non-linear
  quantum properties}} (\bibinfo {year} {2025}),\ \Eprint
  {https://arxiv.org/abs/2505.09206} {arXiv:2505.09206 [quant-ph]} \BibitemShut
  {NoStop}%
\bibitem [{\citenamefont {Chen}\ \emph {et~al.}(2025)\citenamefont {Chen},
  \citenamefont {Wang}, \citenamefont {Yu},\ and\ \citenamefont
  {Zhang}}]{Kean2025}%
  \BibitemOpen
  \bibfield  {author} {\bibinfo {author} {\bibfnamefont {K.}~\bibnamefont
  {Chen}}, \bibinfo {author} {\bibfnamefont {Q.}~\bibnamefont {Wang}}, \bibinfo
  {author} {\bibfnamefont {Z.}~\bibnamefont {Yu}},\ and\ \bibinfo {author}
  {\bibfnamefont {Z.}~\bibnamefont {Zhang}},\ }\href
  {https://arxiv.org/abs/2505.16715} {\bibinfo {title} {Simultaneous estimation
  of nonlinear functionals of a quantum state}} (\bibinfo {year} {2025}),\
  \Eprint {https://arxiv.org/abs/2505.16715} {arXiv:2505.16715 [quant-ph]}
  \BibitemShut {NoStop}%
\bibitem [{\citenamefont {Elben}\ \emph {et~al.}(2023)\citenamefont {Elben},
  \citenamefont {Flammia}, \citenamefont {Huang}, \citenamefont {Kueng},
  \citenamefont {Preskill}, \citenamefont {Vermersch},\ and\ \citenamefont
  {Zoller}}]{Elben2023}%
  \BibitemOpen
  \bibfield  {author} {\bibinfo {author} {\bibfnamefont {A.}~\bibnamefont
  {Elben}}, \bibinfo {author} {\bibfnamefont {S.~T.}\ \bibnamefont {Flammia}},
  \bibinfo {author} {\bibfnamefont {H.-Y.}\ \bibnamefont {Huang}}, \bibinfo
  {author} {\bibfnamefont {R.}~\bibnamefont {Kueng}}, \bibinfo {author}
  {\bibfnamefont {J.}~\bibnamefont {Preskill}}, \bibinfo {author}
  {\bibfnamefont {B.}~\bibnamefont {Vermersch}},\ and\ \bibinfo {author}
  {\bibfnamefont {P.}~\bibnamefont {Zoller}},\ }\href
  {https://doi.org/10.1038/s42254-022-00535-2} {\bibfield  {journal} {\bibinfo
  {journal} {Nature Reviews Physics}\ }\textbf {\bibinfo {volume} {5}},\
  \bibinfo {pages} {9} (\bibinfo {year} {2023})}\BibitemShut {NoStop}%
\bibitem [{\citenamefont {Notarnicola}\ \emph {et~al.}(2023)\citenamefont
  {Notarnicola}, \citenamefont {Elben}, \citenamefont {Lahaye}, \citenamefont
  {Browaeys}, \citenamefont {Montangero},\ and\ \citenamefont
  {Vermersch}}]{Notarnicola_2023}%
  \BibitemOpen
  \bibfield  {author} {\bibinfo {author} {\bibfnamefont {S.}~\bibnamefont
  {Notarnicola}}, \bibinfo {author} {\bibfnamefont {A.}~\bibnamefont {Elben}},
  \bibinfo {author} {\bibfnamefont {T.}~\bibnamefont {Lahaye}}, \bibinfo
  {author} {\bibfnamefont {A.}~\bibnamefont {Browaeys}}, \bibinfo {author}
  {\bibfnamefont {S.}~\bibnamefont {Montangero}},\ and\ \bibinfo {author}
  {\bibfnamefont {B.}~\bibnamefont {Vermersch}},\ }\href
  {https://doi.org/10.1088/1367-2630/acfcd3} {\bibfield  {journal} {\bibinfo
  {journal} {New Journal of Physics}\ }\textbf {\bibinfo {volume} {25}},\
  \bibinfo {pages} {103006} (\bibinfo {year} {2023})}\BibitemShut {NoStop}%
\bibitem [{SMf()}]{SMfootnote1}%
  \BibitemOpen
  \href@noop {} {\bibinfo {title} {Here we assume the space-translational
  symmetry such that the {Green's} function does not depend on
  $(x_1^0+x_2^0)/2$.}}\BibitemShut {Stop}%
\bibitem [{\citenamefont {Littlewood}\ and\ \citenamefont
  {Varma}(1982)}]{Littlewood1982}%
  \BibitemOpen
  \bibfield  {author} {\bibinfo {author} {\bibfnamefont {P.~B.}\ \bibnamefont
  {Littlewood}}\ and\ \bibinfo {author} {\bibfnamefont {C.~M.}\ \bibnamefont
  {Varma}},\ }\href {https://doi.org/10.1103/PhysRevB.26.4883} {\bibfield
  {journal} {\bibinfo  {journal} {Phys. Rev. B}\ }\textbf {\bibinfo {volume}
  {26}},\ \bibinfo {pages} {4883} (\bibinfo {year} {1982})}\BibitemShut
  {NoStop}%
\end{thebibliography}%

\clearpage{}

\onecolumngrid
\appendix
\renewcommand{\thefigure}{S\arabic{figure}}
\setcounter{figure}{0} 
\renewcommand{\thepage}{S\arabic{page}}
\setcounter{page}{1} 
\renewcommand{\theequation}{S.\arabic{equation}}
\setcounter{equation}{0} 
\renewcommand{\thesection}{S\arabic{section}}
\setcounter{section}{0}

\begin{center}
	\large{Supplemental Material for}\\
	\textbf{``Ward-Takahashi Identity and Gauge-Invariant Response Theory for Open Quantum Systems"}
\end{center}

\tableofcontents{}

\section{Derivation of The Ward-Takahashi Identity}

\subsection{Ward-Takahashi Identity in Closed Quantum Systems}

We derive the Ward-Takahashi identity in open quantum systems on the basis of weak U(1) symmetry of the Lindbladian that is used in the main text. To set the basis for our discussion, we first review the derivation of the real-space Ward-Takahashi identity in closed quantum systems in the parlance of path integral. Let $\Psi_{\alpha}(\bm{r},t) :=(c_{\alpha\uparrow}(\bm{r},t),\bar{c}_{\alpha\downarrow}(\bm{r},t))^T$ and $\bar{\Psi}_{\alpha}(\bm{r},t) :=(\bar{c}_{\alpha\uparrow}(\bm{r},t),c_{\alpha\downarrow}(\bm{r},t))$ be Nambu spinors, where $c_{+\sigma}(\bm{r},t)$ and $c_{-\sigma}(\bm{r},t)$ ($\sigma=\uparrow,\downarrow$) are the fermionic fields for the forward and backward paths. The Schwinger-Keldysh action of the system is given by
\begin{equation}
	S=\int_{-\infty}^{\infty}dt\int d\bm{r}(i \bar{\Psi}_+ 
	\partial_t \Psi_+ -H_+- i \bar{\Psi}_- \partial_t \Psi_-+H_-),\label{eq:closed_action}
\end{equation}
where 
\begin{equation}
	H_{\pm}=\frac{1}{2m}(\nabla\bar{\Psi}_{\pm})\tau_3\cdot(\nabla\Psi_{\pm})-\mu\bar{\Psi}_{\pm}\tau_3\Psi_{\pm}+V_{\text{int}}(n_{\pm})
\end{equation}
with $n_{\pm}(\bm{r},t):=\sum_{\sigma}\bar{c}_{\pm\sigma}(\bm{r},t)c_{\pm\sigma}(\bm{r},t)$. This action satisfies the global strong U(1) symmetry, i.e., the action is invariant under the following strong global U(1) transformations:
\begin{equation}\label{eq: strong_U1}
	\Psi_{\alpha}\to e^{i\alpha\theta\tau_3}\Psi_{\alpha},\  \bar{\Psi}_{\alpha}\to\bar{\Psi}_{\alpha}e^{-i\alpha\theta\tau_3},
\end{equation}
where $\theta$ is a constant which is independent of the coordinate. To couple the U(1) gauge field to the action, we employ the substitution $\partial_{\mu}\to\partial_{\mu}-i A_{\alpha\mu}$. Then the action coupled with the gauge field satisfies the local strong U(1) symmetry, i.e., 
\begin{equation}\label{eq: local_strong}
	A_{\alpha\mu}\to A_{\alpha\mu}+\alpha\partial_{\mu}\theta(x),\ \Psi_{\alpha}\to e^{i\alpha\theta(x)\tau_3}\Psi_{\alpha},\  \bar{\Psi}_{\alpha}\to\bar{\Psi}_{\alpha}e^{-i\alpha\theta(x)\tau_3},
\end{equation}
where $x:=(t,\bm{r})$.

We next consider the Green's function defined by
\begin{align}\label{eq: Green-1}
	C_{\alpha \beta} (x_1, x_2) &:= i\frac{1}{Z} \int D [\Psi, \bar{\Psi}]
	\Psi_{\alpha} (x_1) \bar{\Psi}_{\beta} (x_2) e^{i S [A_{\mu}, \Psi,\bar{\Psi}]} \nonumber\\
	&=i\langle\Psi_{\alpha} (x_1) \bar{\Psi}_{\beta} (x_2)\rangle,
\end{align}
where $\langle O\rangle$ is defined as
\begin{equation}
	\langle O\rangle:=\frac{1}{Z}\int D [\Psi, \bar{\Psi}]O e^{i S [A_{\mu}, \Psi,\bar{\Psi}]}
\end{equation}
with the normalization factor
\begin{equation}
	Z:=\int D [\Psi, \bar{\Psi}]e^{i S [A_{\mu}, \Psi,\bar{\Psi}]}.
\end{equation}
We take $A_{+\mu}=-A_{-\mu}=A_{\mu}$ since the vector potential is time-reversal-odd. Under the local strong U(1) transformation, we have
\begin{align}
	C_{\alpha \beta} (x_1, x_2) & =  i \frac{1}{Z}\int D [\Psi, \bar{\Psi}]
	\Psi_{\alpha} (x_1) \bar{\Psi}_{\beta} (x_2) e^{i S [A_{\mu} +
		\partial_{\mu} \theta, e^{i \zeta\theta \tau_3} \Psi_{\zeta},
		\bar{\Psi}_{\zeta} e^{- i \zeta\theta \tau_3}]}\nonumber\\
	& =  i\frac{1}{Z} \int D [\Psi, \bar{\Psi}] e^{- i \alpha\theta (x_1) \tau_3}
	\Psi_{\alpha} (x_1) \bar{\Psi}_{\beta} (x_2) e^{i \beta\theta (x_2) \tau_3} e^{i
		S [A_{\mu} + \partial_{\mu} \theta, \Psi_{\gamma}, \bar{\Psi}_{\gamma}]}\nonumber\\
	& =  C_{\alpha \beta} (x_1, x_2) + \frac{1}{Z}\int D [\Psi, \bar{\Psi}] \int d x \theta (x) \alpha\tau_3 \Psi_{\alpha} (x_1) \bar{\Psi}_{\beta}
	(x_2) \delta
	(x - x_1)  e^{i S [A_{\mu}, \Psi_{\gamma}, \bar{\Psi}_{\gamma}]}\nonumber\\
	& + \frac{1}{Z}\int D [\Psi, \bar{\Psi}] \int d x \theta (x)\left[-\beta\delta (x - x_2) \Psi_{\alpha} (x_1) \bar{\Psi}_{\beta}
	(x_2) \tau_3 + \Psi_{\alpha}
	(x_1) \bar{\Psi}_{\beta} (x_2) \partial_{\mu} \frac{\delta S}{\delta
		A_{\mu}}\right] e^{i S [A_{\mu}, \Psi_{\gamma}, \bar{\Psi}_{\gamma}]} + O
	(\theta^2).
\end{align}
Since the phase $\theta(x)$ is an arbitrary gauge choice, the linear terms in $\theta$ should vanish. This condition gives the Ward-Takahashi identity as
\begin{align}\label{eq: Ward-1}
	&   \pm[\delta (x - x_1) \tau_3 \langle \Psi_{\pm} (x_1) \bar{\Psi}_{\pm}
	(x_2) \rangle - \delta (x - x_2) \langle \Psi_{\pm} (x_1) \bar{\Psi}_{\pm}
	(x_2) \rangle \tau_3] = \langle \Psi_{\pm} (x_1) \bar{\Psi}_{\pm} (x_2)
	\partial_{\mu} J_c^{\mu} \rangle, \\ \label{eq: Ward-2}
	& \pm[\delta (x - x_1) \tau_3 \langle \Psi_{\pm} (x_1) \bar{\Psi}_{\mp}
	(x_2) \rangle + \delta (x - x_2) \langle \Psi_{\pm} (x_1) \bar{\Psi}_{\mp}
	(x_2) \rangle \tau_3] = \langle \Psi_{\pm} (x_1) \bar{\Psi}_{\mp} (x_2)
	\partial_{\mu} J_c^{\mu} \rangle,
\end{align}
where the current $J_c^{\mu}:=-\delta S/\delta A_{\mu}(x)$ gives the full current dressed by the interactions written as
\begin{equation}\label{eq: full}
	J_c^{\mu}(x)=\bar{\Psi}(x)\Gamma^{\mu}(x)\Psi(x)
\end{equation}
with $\Gamma^\mu$ being the full vertex function defined from the current. In a closed quantum system, we usually consider transport properties near the ground state of the Hamiltonian, which is a pure state. Therefore, there is no need to introduce two contours and we only consider the case $\alpha=\beta=+$ in Eq. \eqref{eq: Ward-1}:
\begin{equation}\label{eq: Ward-3}
	\delta (x - x_1) \tau_3 \langle \Psi (x_1) \bar{\Psi}
	(x_2) \rangle - \delta (x - x_2) \langle \Psi (x_1) \bar{\Psi}
	(x_2) \rangle \tau_3 = \langle \Psi (x_1) \bar{\Psi} (x_2)
	\partial_{\mu} J_c^{\mu} \rangle.
\end{equation}
To reproduce the result by Nambu~\cite{Nambu1960}, we assume the spacetime translation symmetry $C_{\alpha\beta}(x_1,x_2)=C_{\alpha\beta}(x_1-x_2)$ and perform the Fourier transformation, obtaining
\begin{align}
	& \int dx_1 dx_2 d x e^{- i (k_1 x_1 + q x) + i k_2 x_2} \delta (x - x_1) \tau_3 \langle
	\Psi(x_1) \bar{\Psi} (x_2) \rangle \nonumber\\
	= & \int dx_1 dx_2  e^{- i (k_1 +q) x_1 + i k_2 x_2} \tau_3 \langle \Psi (x_1) \bar{\Psi}
	(x_2) \rangle \nonumber\\
	= & - i \tau_3 G \left( k_1 + q\right), \label{eq:Fourier-1} \\
	& \int d x_1 d x_2 d x e^{- i (k_1 x_1 + q x) + i
		k_2 x_2} \delta (x - x_2) \langle \Psi (x_1) \bar{\Psi}
	(x_2) \rangle \tau_3 \nonumber\\
	= & \int dx_1 dx_2  e^{- i k_1 x_1
		+ i (k_2 - q) x_2} \langle \Psi (x_1) \bar{\Psi}
	(x_2) \rangle \tau_3 \nonumber\\
	= & - i G \left( k_1 \right)
	\tau_3 , \label{eq:Fourier-2} 
\end{align}
where $G(k)$ is the Green's function in the momentum space.
Here we use $\int dx$ to represent $\int d^{d+1}x$ for convenience. Substituting Eq. \eqref{eq: full} into Eq. \eqref{eq: Ward-3}, we obtain
\begin{equation}\label{eq: Ward-4}
	\langle \Psi (x_1) \bar{\Psi} (x_2)
	\partial_{\mu} (\bar{\Psi}(x) \Gamma^{\mu}
	\Psi (x)) \rangle = -	\partial_{\mu}[ C \left( x_1 - x \right) \Gamma^{\mu} (x)
	C \left( x - x_2 \right)].
\end{equation}
Performing Fourier transformation of Eq. \eqref{eq: Ward-4}, we obtain
\begin{equation}
	\tau_3 G(k + q) - G (k) \tau_3 = G(k) q_{\mu} \Gamma^{\mu} G (k + q),
\end{equation}
which is nothing but the Ward-Takahashi identity in Ref.~\cite{Nambu1960}. 

\subsection{Ward-Takahashi Identity in Open Quantum Systems}\label{sec: open}
We next consider an open quantum system, where the Schwinger-Keldysh action is given by~\cite{Sieberer_2016} 
\begin{equation}\label{eq: SK-open}
	S= \int_{-\infty}^{\infty}dt\int d\bm{r}\left(i \bar{\Psi}_+\partial_t \Psi_+ - H_+- i \bar{\Psi}_- \partial_t \Psi_-+H_{-}+\frac{i\gamma}{2}(\bar{L}_{\tmmathbf{r}+}L_{\tmmathbf{r}+}+\bar{L}_{\tmmathbf{r}-}L_{\tmmathbf{r}-}-2L_{\tmmathbf{r}+}\bar{L}_{\tmmathbf{r}-})\right).
\end{equation}
Here we assume that the Hamiltonian $H$ has U(1) symmetry and the action satisfies the weak U(1) symmetry but breaks the strong U(1) symmetry, which implies that the particle number of the system is not conserved~\cite{Albert2014}. Since the dissipative part in the Lindbladian is local, the dissipative part automatically satisfies the local weak U(1) symmetry. Overall, the action satisfies the global weak U(1) symmetry.

Then we couple the U(1) gauge field to the action \eqref{eq: SK-open} in the same way as in closed quantum systems. The action with the gauge field thus satisfies the local weak U(1) symmetry, i.e.,
\begin{equation}
	A_{\alpha\mu}\to A_{\alpha\mu}+\partial_{\mu}\theta(x),\ \Psi_{\alpha}\to e^{i\theta(x)\tau_3}\Psi_{\alpha},\ \bar{\Psi}_{\alpha}\to\bar{\Psi}_{\alpha}e^{-i\theta(x)\tau_3}.
\end{equation}
We still consider the correlation function \eqref{eq: Green-1} and perform the local strong U(1) transformation \eqref{eq: local_strong} to the action. Since the particle number is not conserved, the action is not invariant under the transformation, which gives rise to
\begin{equation}
	S\to S[\theta]=S-i\gamma\int dtd\bm{r}\theta(x) \left( \frac{\partial \bar{L}_{\bm{r} -}}{\partial \theta(x)}\Big|_{\theta=0} L_{\bm{r} +}
	+  \bar{L}_{\bm{r} -} \frac{\partial L_{\bm{r} +}}{\partial \theta(x)}\Big|_{\theta=0} \right)=:S-i\gamma\int dtd\bm{r}\theta(x)V(x),
\end{equation}
where $\frac{\partial B_{x\alpha}}{\partial \theta(x)}\Big|_{\theta=0}$ describes the first-order derivative of an arbitrary field $B_{x\alpha}$ under the local strong U(1) transformation~\eqref{eq: local_strong} and the additional term containing $V(x):=\frac{\partial \bar{L}_{\bm{r} -}}{\partial \theta(x)}\Big|_{\theta=0} L_{\bm{r} +}
+ \frac{\partial L_{\bm{r} +}}{\partial \theta(x)}\Big|_{\theta=0} \bar{L}_{\bm{r} -} $ describes quantum jumps. For simplicity, we omit the notation $\Big|_{\theta=0}$ in the following. The correlation function transforms as
\begin{eqnarray}
	C_{\alpha \beta} (x_1, x_2) & := & i \frac{1}{Z}\int D [\Psi, \bar{\Psi}]
	\Psi_{\alpha} (x_1) \bar{\Psi}_{\beta} (x_2) e^{i S [A_{\mu} +
		\partial_{\mu} \theta, e^{i \zeta\theta \tau_3} \Psi_{\zeta},
		\bar{\Psi}_{\zeta} e^{- i \zeta\theta \tau_3}]-\gamma\int dtd\bm{r}\theta(x) V(x)}\nonumber\\
	& = & i \frac{1}{Z}\int D [\Psi, \bar{\Psi}] e^{- i \alpha\theta (x_1) \tau_3}
	\Psi_{\alpha} (x_1) \bar{\Psi}_{\beta} (x_2) e^{i \beta\theta (x_2) \tau_3} e^{i
		S [A_{\mu} + \partial_{\mu} \theta, \Psi_{\zeta}, \bar{\Psi}_{\zeta}]-\gamma\int dtd\bm{r}\theta(x) V(x)}\nonumber\\
	& = & C_{\alpha \beta} (x_1, x_2) + \frac{1}{Z}\int D [\Psi, \bar{\Psi}] \int d x \theta (x) [\alpha\tau_3 \delta
	(x - x_1)\Psi_{\alpha} (x_1) \bar{\Psi}_{\beta}
	(x_2) - \beta\delta (x - x_2)\Psi_{\alpha} (x_1) \bar{\Psi}_{\beta}
	(x_2) \tau_3]  e^{i S [A_{\mu}, \Psi_{\zeta}, \bar{\Psi}_{\zeta}]}\nonumber\\
	&  & + \frac{1}{Z}\int D [\Psi, \bar{\Psi}] \int d x \theta (x) \Psi_{\alpha}
	(x_1) \bar{\Psi}_{\beta} (x_2) \partial_{\mu} \frac{\delta S}{\delta
		A_{\mu}} e^{i S [A_{\mu}, \Psi_{\zeta}, \bar{\Psi}_{\zeta}]} \nonumber\\
	& &-i\frac{1}{Z}\int D [\Psi, \bar{\Psi}]\int dx\theta(x) \gamma\Psi_{\alpha}
	(x_1) \bar{\Psi}_{\beta} (x_2) \left( \frac{\partial \bar{L}_{\bm{r} -}}{\partial \theta(x)} L_{\bm{r} +}
	+ \frac{\partial L_{\bm{r} +}}{\partial \theta(x)} \bar{L}_{\bm{r} -} \right) e^{i S [A_{\mu}, \Psi_{\zeta}, \bar{\Psi}_{\zeta}]}+O	(\theta^2).
\end{eqnarray}
The requirement that the terms linear in $\theta$ vanish leads to the following Ward-Takahashi identities:
\begin{align}\label{eq: Ward-open1}
	&   \pm[\delta (x - x_1) \tau_3 \langle \Psi_{\pm} (x_1) \bar{\Psi}_{\pm}
	(x_2) \rangle - \delta (x - x_2) \langle \Psi_{\pm} (x_1) \bar{\Psi}_{\pm}
	(x_2) \rangle \tau_3] = \langle \Psi_{\pm} (x_1) \bar{\Psi}_{\pm} (x_2)
	\partial_{\mu} \bar{J}_c^{\mu} \rangle, \\ \label{eq: Ward-open2}
	& \pm[\delta (x - x_1) \tau_3 \langle \Psi_{\pm} (x_1) \bar{\Psi}_{\mp}
	(x_2) \rangle + \delta (x - x_2) \langle \Psi_{\pm} (x_1) \bar{\Psi}_{\mp}
	(x_2) \rangle \tau_3] = \langle \Psi_{\pm} (x_1) \bar{\Psi}_{\mp} (x_2)
	\partial_{\mu} \bar{J}_c^{\mu} \rangle,
\end{align}
where the full current is defined as
\begin{equation}\label{eq: bar_Jc}
	\bar{J}_c^{\mu}:=J_c^{\mu}+J_d^{\mu}=J_{+}^{\mu}+J_{-}^{\mu}+J_d^{\mu}
\end{equation}
with
\begin{align}\label{eq: jc}
	J^{\mu}_{\alpha}:= -\frac{\delta S}{\delta A_{\alpha\mu}},\  \nabla\cdot\tmmathbf{J}_d = i\gamma \left(
	\bar{L}_{\tmmathbf{r}-} \frac{\partial L_{\tmmathbf{r}+}}{\partial
		\theta(x)} + \frac{\partial \bar{L}_{\tmmathbf{r}-}}{\partial
		\theta(x)} L_{\tmmathbf{r}+} \right),\ J_d^0=0.
\end{align}
We can see that Eq. \eqref{eq: Ward-open1} is nothing but the Ward-Takahashi identity discussed in the main text. The proof is thus completed. 

The dissipative current arises from the equation of continuity, which takes the form of
\begin{align}\label{eq: jcjd}
	\frac{d n_{\bm{r}}}{d t}&=-i[H,n_{\bm{r}}]+\frac{\gamma}{2}([L_{\bm{r}}^{\dagger},n_{\bm{r}}]L_{\bm{r}}+L_{\bm{r}}^{\dagger}[n_{\bm{r}},L_{\bm{r}}])=-i[H,n_{\bm{r}}]+\frac{i\gamma}{2}(\frac{\partial L^{\dagger}_{\bm{r}}}{\partial\theta_{\bm{r}}}L_{\bm{r}}-L_{\bm{r}}^{\dagger}\frac{\partial L_{\bm{r}}}{\partial\theta_{\bm{r}}})  \nonumber\\
	&=	-\nabla\cdot(\bm{j}_c+\bm{j}_d),
\end{align}
where $\bm{j}_c:=i(\Psi^{\dagger} \nabla \Psi - \nabla\Psi^{\dagger} \Psi)/2m$, $\nabla \cdot \tmmathbf{j}_d := \frac{i \gamma}{2} \left(
L_{\tmmathbf{r}}^{\dagger} \frac{\partial L_{\tmmathbf{r}}}{\partial\theta_{\tmmathbf{r}}}|_{\theta=0} - \frac{\partial L_{\tmmathbf{r}}^{\dagger}}{\partial\theta_{\tmmathbf{r}}}|_{\theta=0} L_{\tmmathbf{r}} \right)$, and we define $\partial B_{\bm{r}}/\partial\theta_{\bm{r}}:=i[n_{\bm{r}},B_{\bm{r}}]$ for an arbitrary operator $B_{\bm{r}}$. In Eq. \eqref{eq: jcjd}, the classical current $\bm{j}_c$ corresponds to the current $\bm{J}_c$ in the path integral and the additional current $\bm{j}_d$ corresponds to the dissipative current $\bm{J}_d$ since $L_{\bm{r}\alpha}(\bar{L}_{\bm{r}\alpha})$ follows different transformations for $\alpha=\pm$ as shown in Eq. \eqref{eq: strong_U1}. Therefore, the total current $\bar{J_c}^{\mu}$ is conserved: $\partial_{\mu}\bar{J}_c^{\mu}=0$. This correspondence holds only when the weak U(1) symmetry holds. If the Lindbladian does not satisfy the weak U(1) symmetry, the dissipative current will be
\begin{equation}\label{eq: jd2}
	\nabla\cdot\tmmathbf{J}_d = i\gamma \left(
	\bar{L}_{\tmmathbf{r}-} \frac{\partial L_{\tmmathbf{r}+}}{\partial
		\theta_{x}} + \frac{\partial \bar{L}_{\tmmathbf{r}-}}{\partial
		\theta_{x}} L_{\tmmathbf{r}+} \right)-\frac{i\gamma}{2} \left(
	\frac{\partial \bar{L}_{\tmmathbf{r}+}L_{\tmmathbf{r}+}}{\partial
		\theta_{\bm{r}}} +  \frac{\partial \bar{L}_{\tmmathbf{r}-}L_{\tmmathbf{r}-}}{\partial
		\theta_{\bm{r}}} \right),
\end{equation}
which no longer corresponds to $\bm{j}_d$ since the last term on the right-hand side of Eq. \eqref{eq: jd2} has no counterpart in the equation of continuity. This term represents the non-conservation of the off-diagonal terms in the density matrix. 

\section{Proof of Gauge Invariance}
In this section, we prove the gauge invariance of the response current to an external U(1) gauge field. Following a strategy similar to what is adopted in Ref. \cite{Coleman_2015}, we couple the U(1) gauge field to the system as $\partial_{\mu}\to\partial_{\mu}-iA_{\alpha\mu}$ for the contours $\alpha=\pm$. Since the vector potential is time-reversal-odd, we take $A_{+\mu}=-A_{-\mu}=A_{\mu}$ as our perturbation. Then the response transport current density to the external gauge field is
\begin{equation}
	J_c^{\mu}[A]:=\langle \hat{J}_c^{\mu}[A]\rangle=-\frac{1}{2V}\frac{\partial F}{\partial A_{\mu}},
\end{equation}
where $V$ is the volume of the system and
\begin{equation}
	F:=-i\log Z[A],
\end{equation}
since 
\begin{equation}\label{eq: res}
	-\frac{1}{V}\frac{\partial F}{\partial A_{\mu}}=\frac{i}{V}\frac{1}{Z}\frac{\partial Z}{\partial A_{\mu}}=\frac{1}{V}\frac{1}{Z}\int D\Psi e^{i[S(\Psi)-A_{\mu}\cdot(j^{\mu}_{+}+j^{\mu}_{-})]}(j_{+\mu}[A]+j_{-\mu}[A])=\langle[J_{+\mu}[A]+J_{-\mu}[A]]\rangle=2\langle \hat{J}_{\mu}[A]\rangle.
\end{equation}
Here we use $(j_{+\mu}[A]+j_{-\mu}[A])/2$ to represent the classical component of the current operator, whose average corresponds to the current which can be observed experimentally. If we assume the London equation
\begin{equation}
	J_c^{\mu}=-K^{\mu\nu}A_{\nu},
\end{equation}
the superfluid-weight matrix $K^{\mu\nu}$ can be obtained from
\begin{equation}
	K^{\mu\nu}=\frac{1}{2V}\frac{\partial^{2}F}{\partial A_{\mu}\partial A_{\nu}}\Big|_{A=0}.\label{Av}
\end{equation}
The form of perturbation in the action contains linear terms and quadratic terms. The latter contribute to the Feynman diagram (a) in Fig. 1 in the main text and the former contribute to the Feynman diagram (b) in Fig. 1 in the main text. Under the gauge transformation, the current change $\delta J_1$ from the Feynman diagram (a) takes the form of
\begin{equation}
	\delta J_1^{\mu} (x) = \delta_{\mu \nu} (1 - \delta_{\mu 0}) \frac{1}{2
		m} \text{Tr} [\tau_3 \langle \Psi_+ (x) \bar{\Psi}_+ (x) - \Psi_- (x)
	\bar{\Psi}_- (x) \rangle] \partial_{\nu} \phi (x),
\end{equation}
where we consider the current in the space-time coordinates since the Lindblad dynamics does not have time-translational symmetry. By defining non-equilibrium Green's functions~\cite{SMfootnote1}
\begin{align}
	G^T (k, t) & := - i \int d (x_1 - x_2) \text{Tr} \langle \Psi_+
	(x_1) \bar{\Psi}_+ (x_2) \rangle e^{i k \cdot (x_1 - x_2)} ,\\
	G^{\tilde{T}} (k, t) &:= - i \int d (x_1 - x_2) \text{Tr} \langle \Psi_-
	(x_1) \bar{\Psi}_- (x_2) \rangle e^{i k \cdot (x_1 - x_2)} ,
\end{align}  
where $t:=(x^0_1+x^0_2)/2$, we can rewrite the perturbed current as
\begin{equation}
	\delta J_1^{\mu} (x) =	 i \delta_{\mu \nu} (1 - \delta_{\mu 0}) \frac{1}{2 m} \int \frac{d^{d+1}k}{(2\pi)^{d+1}}
	\text{Tr}\left[ \tau_3 (G^T (k, t) - G^{\tilde{T}} (k, t))\right] \partial_{\nu} \phi
	(x).
\end{equation}
Below we use $\int dk$ to represent $\int d^{d+1}k/(2\pi)^{d+1}$ for convenience. Here we note that the Green's function depends not only on the momentum and energy but also on the averaged time $t$. Then using
\begin{equation}\label{eq: def_n}
	n (t) := - \frac{1}{2} \int dk \text{Tr} [\tau_3 (i G^T (k, t) - i
	G^{\tilde{T}} (k, t))] = - \frac{1}{2} \int d k \text{Tr} [\tau_3 (i G^R
	(k, t) + i G^A (k, t))], 
\end{equation}
the contribution from the first diagram is expressed as
\begin{equation}
	\delta J_1^{\mu} (x) = - \frac{n (t)}{m} (1 - \delta_{\mu 0}) \partial_{\mu}
	\phi (x) . \label{eq:J1}
\end{equation}
In deriving the second equality in Eq. \eqref{eq: def_n}, we use the relations~\cite{Kamenev_2011}
\begin{equation}
	G^R=G^T-G^<,\ G^A=G^<-G^{\tilde{T}},
\end{equation}
where the lesser Green's function is defined as
\begin{equation}
	G^< (k, t) := - i \int d (x_1 - x_2) \text{Tr} \langle \Psi_+
	(x_1) \bar{\Psi}_- (x_2) \rangle e^{i k \cdot (x_1 - x_2)} .
\end{equation}

We next examine the Feynman diagram (b) in Fig. 1 in the main text. The perturbed current $\delta J_2$ can be written as
\begin{align}\label{eq: J2_2}
	\delta J_2^{\mu} (x) & = \frac{i}{2} \sum_{\alpha} \int dx' \Tr\langle \gamma_{\alpha}^{\nu} \Psi_{\alpha} (x') J_c^{\mu} (x)\bar{\Psi}_{\alpha} (x') \rangle \partial_{\nu}' \phi (x')\nonumber\\
	& = \frac{i}{2} \sum_{\alpha} \int dx' \Tr\langle \gamma_{\alpha}^{\nu} \Psi_{\alpha} (x') \bar{J}_c^{\mu} (x)\bar{\Psi}_{\alpha} (x') \rangle \partial_{\nu}' \phi (x'),
\end{align}
where $\partial_{\nu}' {:= \partial / \partial x'}^{\nu}$ and the factor $1/2$ originates from the average over the contours. The second equality in Eq. \eqref{eq: J2_2} holds because the dissipative current $\bm{j}_d$ is decoupled from the EM field and therefore does not response to it. Hence, we can replace $J_c^{\mu}$ with $\bar{J}_c^{\mu}=J_c^{\mu}+J_d^{\mu}$ in Eq. \eqref{eq: J2_2}. Furthermore, Eq. \eqref{eq: J2_2} can be simplified as
\begin{eqnarray}
	\delta J_2^{\mu} (x) & = & \frac{i}{2}\sum_{\alpha} \int dx' \Tr \langle
	\gamma_{\alpha}^{\mu} \Psi_{\alpha} (x) \bar{J}_c^{\nu} (x') \bar{\Psi}_{\alpha} (x)
	\rangle \partial_{\nu}' \phi (x') \nonumber\\
	& = & - \frac{i}{2}\sum_{\alpha} \int dx' \Tr \langle \gamma_{\alpha}^{\mu}
	\Psi_{\alpha} (x) \partial_{\nu}' \bar{J}_c^{\nu} (x') \bar{\Psi}_{\alpha} (x)
	\rangle \phi (x') \nonumber\\
	& = & -\frac{i}{2} \lim_{x_1 \rightarrow x_2} \sum_{\alpha}\alpha \int dx' (\delta (x' -
	x_1) - \delta (x' - x_2)) \Tr \langle \gamma_{\alpha}^{\mu} \tau_3
	\Psi_{\alpha} (x_1) \bar{\Psi}_{\alpha} (x_2) \rangle \phi (x'),
\end{eqnarray} 
where $x_2=x$ and we have substituted the Ward-Takahashi identity \eqref{eq: Ward-open1} in obtaining the third equality. In the first line, we have used the fact that $\delta J_2^{\mu}$ in Eq. (\ref{eq: J2_2}) can be written as
\begin{equation}\label{eq: J12}
	\delta\mathcal{J}_2^{\mu}= \frac{i}{2} \sum_{\alpha} \int dx' \Tr\langle \gamma_{\alpha}^{\nu} \Psi_{\alpha} (x) \bar{J}_c^{\mu} (x')\bar{\Psi}_{\alpha} (x) \rangle \partial_{\nu}' \phi (x').
\end{equation}
This replacement is shown as follows. First, $\partial_{\mu}\delta\mathcal{J}_2^{\mu}=\partial_{\mu}\delta J_2^{\mu}$ since the both sides are proportional to $\delta(x-x')$ due to the Ward-Takahashi identity (\ref{eq: Ward-open1}). Then we have $\delta\mathcal{J}_2^{\mu}=\delta J_2^{\mu}+C$, where $C$ is a constant. Since the two currents should both vanish when $\phi=0$, we obtain $C=0$.

By performing the Fourier transformation to the relative coordinates, we obtain
\begin{align}
	\delta (x' - x_1) \text{Tr} \langle \gamma_{\alpha}^{\mu} \tau_3
	\Psi_{\alpha} (x_1) \bar{\Psi}_{\alpha} (x_2) \rangle&=-i \delta (x' - x_1) \text{Tr} \left[ \gamma_{\alpha}^{\mu} \tau_3
	C_{\alpha\alpha} \left( x_1 - x_2, \frac{x_1^0 + x_2^0}{2} \right) \right]\nonumber\\
	&= -i \int d q' d k_1 e^{- i q' \cdot (x_1 - x')} \text{Tr} \left[
	\gamma_{\alpha}^{\mu} \tau_3 G_{\alpha\alpha} \left( k_1, \frac{x_1^0 + x_2^0}{2}
	\right) \right] e^{i k_1 \cdot (x_1 - x_2)},\\
	\delta (x' - x_2) \text{Tr} \langle \gamma_{\alpha}^{\mu} \tau_3
	\Psi_{\alpha} (x_1) \bar{\Psi}_{\alpha} (x_2) \rangle = & -i \delta (x' - x_2) \text{Tr} \left[ \gamma_{\alpha}^{\mu} \tau_3
	C_{\alpha\alpha} \left( x_1 - x_2, \frac{x_1^0 + x_2^0}{2} \right) \right]\nonumber\\
	&= -i \int d q' d k_1 e^{- i q' \cdot (x_2 - x')} \text{Tr} \left[
	\gamma_{\alpha}^{\mu} \tau_3 G_{\alpha\alpha} \left( k_1, \frac{x_1^0 + x_2^0}{2}
	\right) \right] e^{i k_1 \cdot (x_1 - x_2)} . 
\end{align}
Since the bare vertex operators take the form of
\begin{equation}
	\gamma^i_{\alpha} := \frac{i }{2 m} \left(
	\frac{\partial}{\partial x_1^i} - \frac{\partial}{\partial x_2^i} \right),
	\gamma_{\alpha}^0 =  \tau_3,
\end{equation}
the perturbed current $\delta J_2$ can be written as
\begin{eqnarray}
	\delta J_2^{\mu} (x) & = & \frac{i}{2} \sum_{\alpha} \alpha \int d x' \int d q' d k_1
	\frac{{q'}^{\mu}}{m} \text{Tr} [\tau_3 iG_{\alpha\alpha} (k_1, t)] e^{- i q' \cdot (x
		- x')} \phi (x') \nonumber\\
	& = & - \frac{n (t)}{m} \int d x' \int d q' \left( {i q'}^{\mu} \right)
	e^{- i q' \cdot (x - x')} \phi (x') \nonumber\\
	& = & \frac{n (t)}{m} \int d x' \int d q' \partial_{\mu} e^{- i q' \cdot (x
		- x')} \phi (x') \nonumber\\
	& = & \frac{n (t)}{m} \int d x' \partial_{\mu} \delta (x - x') \phi (x')
	\nonumber\\
	& = & \frac{n (t)}{m} \partial_{\mu} \phi (x), 
\end{eqnarray}
for $\mu=1,2,3$ and $\delta J_2(x)=0$ for $\mu=0$. Therefore, we can rewrite $\delta J_2$ as
\begin{equation}
	\delta J_2^{\mu} = \frac{n (t)}{m} (1 - \delta_{\mu 0}) \partial_{\mu} \phi
	(x) . \label{eq:J2}
\end{equation}
Combining Eq. \eqref{eq:J1} with Eq. \eqref{eq:J2}, we obtain the correction to the
current as
\begin{equation}
	\delta J^{\mu} = \delta J_1^{\mu} + \delta J_2^{\mu} = 0.
\end{equation}
Thus the gauge invariance is proved.

Since the current is gauge-invariant, we can write down the expression for $\delta J^{\mu}$ as
\begin{equation}
	\delta J^{\mu} (\tmmathbf{q}, t) = - \frac{n (t)}{m} (1 - \delta_{\mu 0})
	\left( \delta^{\mu j} - \frac{q^{\mu} q^j}{| \tmmathbf{q} |^2} \right) A_j
	(\tmmathbf{q}, t), \label{eq:response1}
\end{equation}
if we take the Hamilton gauge, i.e., $A_0 = 0$, which satisfies the conservation law: $\partial_{\mu} \delta J^{\mu} (\tmmathbf{r}, t) = 0$. In this case, we have the gauge invariance under $A_{\mu} \rightarrow A_{\mu}
+ \partial_{\mu} \phi (\tmmathbf{x})$ for Eq. \eqref{eq:response1}.

We emphasize that the gauge invariance shown above can be regarded as a consequence of the $f$-sum rule derived in Ref. \cite{Hongchao2024}. To see this, we first write the $f$-sum rule in open quantum systems in the presence of weak U(1) symmetry as (see Eq. (81) in the Supplemental Material of Ref. \cite{Hongchao2024})
\begin{equation}\label{eq: f-sum}
	N(t_0)=\frac{k_{i}k_{j}}{k^{2}}\int\frac{d\omega_{1}}{2\pi\omega_{1}}\gamma^{i,j}(\bm{k},\omega_{1},t_{0}),
\end{equation}
where $N(t_0)$ is the particle number of the system at an arbitrarily chosen time $t_0$ and 
\begin{equation}
	\gamma^{i,j}(\bm{k},\omega,t_{0})=m\int dte^{i\omega t}\langle[j_t^{i}(\bm{k},t+t_{0}),j_t^{j}(-\bm{k},t_{0})]\rangle=m\langle[j_t^{i}(\bm{k},\omega),j_t^{j}(-\bm{k},t_{0})]\rangle e^{-i\omega t_{0}}
\end{equation}
is the current-current correlation function with $\bm{j}_t:=\bm{j}_c+\bm{j}_d$ being the total current including both the transport current and the dissipative current. When we couple the EM field to the Hamiltonian, there are two types of current: the paramagnetic current $\delta \bm{J}_{\rm{pm}}$ originating from the linear coupling term $\mathcal{O}(A)$ and the diamagnetic current $\delta \bm{J}_{\rm{dm}}$ from the quadratic coupling term $\mathcal{O}(A^2)$. When we perturb the Hamiltonian by $H\to H-\int d\bm{r}\bm{A}\cdot\bm{j}$, we obtain the paramagnetic current from Eq. (86) in the Supplemental Material of Ref. \cite{Hongchao2024} as
\begin{equation}\label{eq: pm1}
	\langle\delta J^i_{\rm{pm}}(\bm{k},t)\rangle=\int\frac{d\omega_{1}}{2\pi\omega_{1}}\int d^{d}\bm{r'}e^{-i\omega_{1}t}\langle[j_t^i(\bm{k},t),j_t^j(\bm{r}',\omega_{1})]\rangle A_{j}(\bm{r}').
\end{equation}
Here we have assumed that the frequency of the EM field is small such that the EM field can be considered to be independent of time and that the momentum of the EM field is $\bm{k}$. By performing Fourier transformation in Eq. \eqref{eq: pm1}, we have
\begin{equation}\label{eq: pm}
	\langle\delta J^i_{\rm{pm}}(\bm{k},t)\rangle=\frac{1}{V}\int\frac{d\omega_{1}}{2\pi\omega_{1}}e^{-i\omega_{1}t}\langle[j_t^i(\bm{k},t),j_t^j(\bm{-k},\omega_{1})]\rangle A_{j}(\bm{k}).
\end{equation}
In the longitudinal gauge $\nabla\times\bm{A}=0$, we can rewrite the EM field as $A_j=a_{\bm{k}}k_j$. Then by substituting Eq. \eqref{eq: f-sum} into Eq. \eqref{eq: pm}, we obtain
\begin{align}\label{eq: f-sum2}
	k_i\langle\delta J^i_{\rm{pm}}(\bm{k},t)\rangle&=k_ik_j\frac{1}{V}\int\frac{d\omega_{1}}{2\pi\omega_{1}}e^{-i\omega_{1}t}\langle[j_t^i(\bm{k},t),j_t^j(\bm{-k},\omega_{1})]\rangle a_{\bm{k}}\nonumber\\ 
	&=\frac{N(t)|\bm{k}|^2}{mV}a_{\bm{k}}.
\end{align}
On the other hand, the diamagnetic current $\delta J_{\rm{dm}}$ is given by~\cite{Anderson1958_SC}
\begin{equation}
	\langle\delta J^i_{\rm{dm}}(\bm{k},t)\rangle=-\frac{N(t)}{mV}A^i(\bm{k})=-\frac{N(t)}{mV}k^ia_{\bm{k}}.
\end{equation}
Therefore, the divergence of the diamagnetic current takes the form of
\begin{equation}
	k_i\langle\delta J^i_{\rm{dm}}(\bm{k},t)\rangle=-\frac{N(t)}{mV}k_iA^i(\bm{k})=-\frac{N(t)}{mV}|\bm{k}|^2a_{\bm{k}}.
\end{equation}
Together with Eq. \eqref{eq: f-sum2}, we have
\begin{equation}
	k_i\langle\delta J^i\rangle=	k_i\langle\delta J^i_{\rm{pm}}(\bm{k},t)+\delta J^i_{\rm{dm}}(\bm{k},t)\rangle=0,
\end{equation}
which implies that the longitudinal component of the response current vanishes if the EM field is longitudinal. Hence, the gauge invariance can be considered as a consequence of the $f$-sum rule.

The physical meaning of $n(t)$ in Eq. \eqref{eq:response1} depends on the specific form of the Hamiltonian and Lindblad operators. Here we consider a three-dimensional dissipative superconductor under two-body loss as an example. The Schwinger-Keldysh action is given by Eq. \eqref{eq: SK-open} with the Hamiltonian
\begin{equation}
	H_{\alpha}=\sum_{\bm{k},\sigma}\varepsilon_{\bm{k}}\bar{c}_{\bm{k}\sigma\alpha}c_{\bm{k}\sigma\alpha}-U\int d\bm{r}\bar{c}_{\bm{r}\uparrow\alpha}\bar{c}_{\bm{r}\downarrow\alpha}c_{\bm{r}\downarrow\alpha}c_{\bm{r}\uparrow\alpha},
\end{equation}
where $U>0$ is the strength of an attractive interaction and the Lindblad operator for two-body loss is taken to be $L_{\bm{r}\alpha}=c_{\bm{r}\downarrow\alpha}c_{\bm{r}\uparrow\alpha}$~\cite{Yamamoto2021}. Employing the mean-field approximation to the action, we obtain the Schwinger-Keldysh effective action as~\cite{Yamamoto2021}
\begin{equation}\label{eq: BCS_action}
	S=\int_{-\infty}^{\infty}\sum_{\bm{k}}\left[\bar{\Psi}_{+}\left(\begin{array}{cc}
		i\partial_t-\varepsilon_{\bm{k}} & -\Delta\\
		-\Delta^{\ast} & -i\partial_t+\varepsilon_{\tmmathbf{k}}
	\end{array}\right)\Psi_{+}-\bar{\Psi}_{-}\left(\begin{array}{cc}
		i\partial_t-\varepsilon_{\bm{k}} & -\Delta\\
		-\Delta^{\ast} & -i\partial_t+\varepsilon_{\tmmathbf{k}}
	\end{array}\right)\Psi_{-}\right],
\end{equation}
with $\Delta$ defined as
\begin{equation}
	\Delta:=-\frac{U_c}{N}\sum_{\bm{k}}\bra c_{-\bm{k}\downarrow}c_{\bm{k}\uparrow}\ket,
\end{equation}
where $U_c:=U+i\gamma/2$ is complex-valued.
We further make the quasi-steady-state approximation, under which the density $n$ can be considered as a constant over a long time period, and obtain
\begin{eqnarray}\label{eq: inte}
	\int \frac{d^4 k}{(2 \pi)^4} \text{Tr} [\tau_3 G^R (k)] & = & \int \frac{d^4
		k}{(2 \pi)^4} \frac{1}{(k^0 + i 0^+)^2 - E_{\tmmathbf{k}}^2} \text{Tr}
	\left[ \tau_3 \left(\begin{array}{cc}
		k^0 + \varepsilon_{\tmmathbf{k}} & \Delta\\
		\Delta^{\ast} & k^0 - \varepsilon_{\tmmathbf{k}}
	\end{array}\right) \right] \nonumber\\
	& = & \int \frac{d^3 k}{(2 \pi)^3} \frac{d k^0}{2 \pi} \frac{1}{(k^0 + i
		0^+)^2 - E_{\bm{k}}^2} \text{Tr} \left[ \tau_3 \left(\begin{array}{cc}
		(k^0 + \varepsilon_{\bm{k}}) e^{i 0^{+} k^0} & \Delta\\
		\Delta^{\ast} & (k^0 - \varepsilon_{\bm{k}}) e^{- i 0^+ k^0}
	\end{array}\right) \right] \nonumber\\
	& = & i \int \frac{d^3 k}{(2 \pi)^3} 1 \nonumber\\
	& = & i N / V\nonumber\\
	& = & i n,
\end{eqnarray}
where $n$ denotes the total fermion density and $E_{\bm{k}}:=\sqrt{\varepsilon_{\bm{k}}^2+|\Delta|^2}$. Similarly, the integration over the advanced Green's function gives the same result. Therefore,
\begin{equation}
	- \frac{1}{2} \int \frac{d^4 k}{(2 \pi)^4} \text{Tr} [\tau_3 (i G^R (k) + i
	G^A (k))] = n.
\end{equation}
Hence, $n(t)$ in Eq. (\ref{eq:response1}) for this case represents the total fermion density at time $t$. Due to the particle loss, the value of $n(t)$ decays in time. 

The above analysis neglects the one-body-loss channel~\cite{Mazza2023}. If we take it into account, the integral in Eq. \eqref{eq: inte} becomes
\begin{eqnarray}
	- \int \frac{d^4 k}{(2 \pi)^4} \text{Tr} [\tau_3 i G^R (k)] & = & - i \int
	\frac{d^4 k}{(2 \pi)^4} \frac{1}{\left( k^0 + i \frac{\gamma n}{2} \right)^2
		- E_{\bm{k}}^2} \text{Tr} \left[ \tau_3 \left(\begin{array}{cc}
		k^0 + \varepsilon_{\bm{k}} + \frac{i \gamma n}{2} & \Delta\\
		\Delta^{\ast} & k^0 - \varepsilon_{\bm{k}} + \frac{i \gamma n}{2}
	\end{array}\right) \right] \nonumber\\
	& = & \int \frac{d^3 k}{(2 \pi)^3} 1 \nonumber\\
	& = & N (t) / V\nonumber\\
	& = &  n (t),
\end{eqnarray}
where the retarded Green's function is given by
\begin{equation}
	G^R(k,t)=\frac{1}{\left( k^0 + i \frac{\gamma n}{2} \right)^2
		- E_{\bm{k}}^2} \left(\begin{array}{cc}
		k^0 + \varepsilon_{\bm{k}} + \frac{i \gamma n}{2} & \Delta\\
		\Delta^{\ast} & k^0 - \varepsilon_{\bm{k}} + \frac{i \gamma n}{2}
	\end{array}\right).
\end{equation}
Hence, the one-body-loss channel does not influence the final result.
\section{Nambu-Goldstone Mode}
In this section, we derive the Nambu-Goldstone (NG) mode which arises as a consequence of weak U(1) symmetry breaking. Since the NG mode depends on the type of dissipation, here we consider the three-dimensional dissipative superconductivity under the two-body loss. The action is given by Eq. \eqref{eq: SK-open} through replacement of the Lindblad operator with $L_{\bm{r}\alpha}=c_{\alpha\downarrow}(x)c_{\alpha\uparrow}(x)$ where we use $x:=(t,\bm{r})$. Under the mean-field approximation, we obtain
\begin{equation}\label{eq: action_MF}
	S = \int_{- \infty}^{\infty} d \omega\int d^3 \tmmathbf{k} \bar{\Psi}_o  \left(\begin{array}{cccc}
		\omega - \varepsilon_{\tmmathbf{k}} + \frac{i \gamma n}{2} & - \Delta & 0
		& 0\\
		- \Delta^{\ast} & \omega + \varepsilon_{\tmmathbf{k}} - \frac{i \gamma
			n}{2} & 0 & i \gamma n\\
		- i \gamma n & 0 & - \omega + \varepsilon_{\tmmathbf{k}} + \frac{i \gamma
			n}{2} & \Delta\\
		0 & 0 & \Delta^{\ast} & - \omega - \varepsilon_{\tmmathbf{k}} - \frac{i
			\gamma n}{2}
	\end{array}\right) 
	\Psi_o,
\end{equation}
where $\omega$ plays the same role as $k^0$ above and we introduce the generalized Nambu spinor on the Keldysh contour as
\begin{equation}
	\bar{\Psi}_o = (\bar{c}_{k \uparrow +}, c_{k \downarrow +}, \bar{c}_{k
		\uparrow -}, c_{k \downarrow -}), \Psi_o = \left(\begin{array}{c}
		c_{k \uparrow +}\\
		\bar{c}_{k \downarrow +}\\
		c_{k \uparrow -}\\
		\bar{c}_{k \downarrow -}
	\end{array}\right).
\end{equation}
Here we take into account not only the BCS-pairing terms but also the Hartree-Fock terms. The approximation takes the one-body-loss channel into account~\cite{Mazza2023}. We note that both the order parameter $\Delta$ and the total fermion density $n$ are time-dependent. For simplicity, we employ the quasi-steady-state approximation, where the parameters change slowly and remain nearly constant over a long time. We can find the Green's functions from the action as
\begin{eqnarray}
	G^R & = & G^T - G^{<} \nonumber\\
	& = & \frac{1}{\omega^2 - (\varepsilon_{\tmmathbf{k}}^2 + | \Delta |^2) + i
		\omega \gamma n - \gamma^2 n^2 / 4} \left(\begin{array}{cc}
		\omega + \varepsilon_{\tmmathbf{k}} + \frac{i \gamma n}{2} & \Delta\\
		\Delta^{\ast} & \omega - \varepsilon_{\tmmathbf{k}} + \frac{i \gamma n}{2}
	\end{array}\right), \label{eq:R} \\
	G^A & = & (G^R)^{\dagger} \nonumber\\
	& = & \frac{1}{\omega^2 - (\varepsilon_{\tmmathbf{k}}^2 + | \Delta |^2) - i
		\omega \gamma n - \gamma^2 n^2 / 4} \left(\begin{array}{cc}
		\omega + \varepsilon_{\tmmathbf{k}} - \frac{i \gamma n}{2} & \Delta\\
		\Delta^{\ast} & \omega - \varepsilon_{\tmmathbf{k}} - \frac{i \gamma n}{2}
	\end{array}\right), \label{eq:A} \\
	G^{<} & = & \frac{1}{[(\omega^2 - (\varepsilon_{\tmmathbf{k}}^2 + | \Delta
		|^2) - \gamma^2 n^2 / 4)^2 + \omega^2 \gamma^2 n^2]} \left(\begin{array}{cc}
		i \gamma n | \Delta |^2 & i \gamma n \Delta (\varepsilon_{\tmmathbf{k}} -
		\omega)\\
		i \gamma n \Delta^{\ast} (\varepsilon_{\tmmathbf{k}} - \omega) & i \gamma
		n (\varepsilon_{\tmmathbf{k}} - \omega)^2
	\end{array}\right), \\
	G^T & = & \frac{1}{A (\omega)} \left(\begin{array}{cc}
		((\omega + \varepsilon_{\tmmathbf{k}}) - i \gamma n / 2) (\omega^2 -
		E_{\tmmathbf{k}}^2 - i \gamma n \varepsilon_{\tmmathbf{k}}) & \Delta (i \gamma n
		\varepsilon_{\tmmathbf{k}} - (\omega^2 - E_{\tmmathbf{k}}^2))\\
		\Delta (i \gamma n \varepsilon_{\tmmathbf{k}} - (\omega^2 -
		E_{\tmmathbf{k}}^2)) & ((\omega - \varepsilon_{\tmmathbf{k}}) + i \gamma
		n / 2) (\omega^2 - E_{\tmmathbf{k}}^2 - i \gamma n \varepsilon_{\tmmathbf{k}})
	\end{array}\right),
\end{eqnarray}
where $A(\omega):=(\omega^2 - (\varepsilon_{\tmmathbf{k}}^2 + | \Delta
|^2) - \gamma^2 n^2 / 4)^2 + \omega^2 \gamma^2 n^2$. We now derive the NG mode from the vertex $\Gamma^{\mu}$. In Sec. \ref{sec: open} we consider the Ward-Takahashi identity for the vertex of the classical component of the current~\cite{Sieberer_2016} by performing the strong U(1) transformation. 
%In addition, we can also perform the weak U(1) transformation to the action \eqref{eq: SK-open} by coupling the EM field to it. 
Following a similar procedure, we perform the local weak U(1) transformation to obtain the Ward-Takahashi identity for the vertex of the quantum component of the current as 
\begin{align}\label{eq: Ward-open3}
	[\delta (x - x_1) \tau_3 \langle \Psi_{\alpha} (x_1) \bar{\Psi}_{\beta}
	(x_2) \rangle - \delta (x - x_2) \langle \Psi_{\alpha} (x_1) \bar{\Psi}_{\beta}
	(x_2) \rangle \tau_3] = \langle \Psi_{\alpha} (x_1) \bar{\Psi}_{\beta} (x_2)
	\partial_{\mu} J_q^{\mu} \rangle, 
\end{align}
where the quantum component of the current is given by $J_q^{\mu}=J_+^{\mu}-J_-^{\mu}$. Under the quasi-steady-state approximation, we can perform Fourier transformation to Eq. \eqref{eq: Ward-open3}, obtaining
\begin{equation}
	\tau_3 G_{\alpha \beta} (k + q) - G_{\alpha \beta} (k) \tau_3 = G_{\alpha
		\gamma} (k) q_{\mu} \Gamma_{q,\gamma \sigma}^{\mu} G_{\sigma \beta} (k + q),
\end{equation}
where $\Gamma_{q}^{\mu}$ is the vertex for the quantum component $J_q^{\mu}$ of the current. Then we take the retarded component of both sides as
\begin{equation}
	\tau_3 G^R (k + q) - G^R (k) \tau_3 = G^R (k) q_{\mu} \Gamma_{q, R}^{\mu}
	G^R (k + q). \label{eq:Ward-R}
\end{equation}
Hence, the retarded quantum vertex is given by
\begin{equation}\label{eq: retarded}
	q_{\mu} \Gamma_{q, R}^{\mu} = (G^R (k))^{- 1} \tau_3 - \tau_3 (G^R (k +
	q))^{- 1}.
\end{equation}
By substituting Eq. \eqref{eq:R} into Eq. \eqref{eq: retarded}, we obtain
\begin{eqnarray}\label{eq: retarded_vertex}
	q_{\mu} \Gamma^{\mu}_{q,R} & = & (G^R (k))^{- 1} \tau_3 - \tau_3 (G^R (k +
	q))^{- 1} \nonumber\\
	& = & \left(\begin{array}{cc}
		\omega - \varepsilon_{\tmmathbf{k}} + \frac{i \gamma n}{2} & \Delta\\
		\Delta & \omega + \varepsilon_{\tmmathbf{k}} + \frac{i \gamma n}{2}
	\end{array}\right) \tau_3 - \tau_3 \left(\begin{array}{cc}
		\omega - \varepsilon_{\tmmathbf{k}+\tmmathbf{q}} + \frac{i \gamma n}{2} &
		\Delta\\
		\Delta & \omega + \varepsilon_{\tmmathbf{k}+\tmmathbf{q}} + \frac{i
			\gamma n}{2}
	\end{array}\right) \nonumber\\
	& = & - 2 i \Delta \tau_2 + (\varepsilon_{\tmmathbf{k}+\tmmathbf{q}} -
	\varepsilon_{\tmmathbf{k}}) \tau_0 . 
\end{eqnarray}
Here we absorb the phase of the order parameter into the EM field. When we take the limit $q\to0$, the right-hand side of Eq. \eqref{eq: retarded_vertex} remains finite while the left-hand side vanishes. Therefore, there must exist a singularity in $\Gamma_{q,R}$ for low-energy modes. In the following we discuss the dispersion relation of the NG mode. According to Ref. \cite{Nambu1960}, we can write the recursion relation of the full vertex as
\begin{equation}\label{eq: recursion}
	\Gamma_{q,\alpha \delta}^{\mu} = \gamma_{\alpha \delta}^{\mu} + i \sum_{\beta,\gamma}\int
	\frac{d^4 k}{(2 \pi)^4} \tau_3 G_{\delta \beta} (k + q) \Gamma_{q,\beta
		\gamma}^{\mu} (k + q, k) G_{\gamma \alpha} (k) \tau_3 V_{\alpha \delta} (p -
	k) .
\end{equation}
From the effective action, we have the interaction lines as
\begin{equation}
	V_{+ +} = U - i \gamma / 2, V_{- -} = - (U + i \gamma / 2), V_{- +} = i
	\gamma .
\end{equation}
To obtain the NG mode~\cite{Littlewood1982}, we use $\Gamma_{q,R}^0(q)=\phi(q)\tau_2$ to rewrite Eq. \eqref{eq: recursion} as
\begin{equation}
	\tau_2 = i \int \frac{d^4 k}{(2 \pi)^4} (U - i \gamma / 2) \tau_3 [G^T (k +
	q) \tau_2 G^T (k)] \tau_3, \label{eq:corrected}
\end{equation}
which is valid up to the correction of $O(\gamma^2)$. We first consider the zeroth-order relation ($O(\gamma^0)$) given by
\begin{equation}
	1 + i U \int \frac{d^3 k}{(2 \pi)^3} \int \frac{d k^0}{2 \pi}
	\frac{- k^0 (k^0 + q^0) + \varepsilon_{\tmmathbf{k}}
		\varepsilon_{\tmmathbf{k}+\tmmathbf{q}} + \Delta^2}{(k^0 + q^0 +
		E_{\tmmathbf{k}+\tmmathbf{q}}) (k^0 + q^0 - E_{\tmmathbf{k}+\tmmathbf{q}})
		(k^0 - E_{\tmmathbf{k}}) (k^0 + E_{\tmmathbf{k}})} = 0.
\end{equation}
After performing the integration over the energy, we have
\begin{equation}\label{eq: gap2}
	1 + \frac{1}{2} U \int \frac{d^3 k}{(2 \pi)^3} \frac{E_{\tmmathbf{k}} +
		E_{\tmmathbf{k}+\tmmathbf{q}}}{E_{\tmmathbf{k}}
		E_{\tmmathbf{k}+\tmmathbf{q}}} \frac{E_{\tmmathbf{k}}
		E_{\tmmathbf{k}+\tmmathbf{q}} + \varepsilon_{\tmmathbf{k}}
		\varepsilon_{\tmmathbf{k}+\tmmathbf{q}} + \Delta^2}{(q^0)^2 -
		(E_{\tmmathbf{k}} + E_{\tmmathbf{k}+\tmmathbf{q}})^2} = 0.
\end{equation}
Substituting the gap equation
\begin{equation}
	1=U\int\frac{d^3k}{(2\pi)^3}\frac{1}{2E_{\bm{k}}}
\end{equation}
in Eq. \eqref{eq: gap2}, we have
\begin{equation}
	\frac{1}{4} U \int \frac{d^3 k}{(2 \pi)^3} \frac{E_{\tmmathbf{k}} +
		E_{\tmmathbf{k}+\tmmathbf{q}}}{E_{\tmmathbf{k}}
		E_{\tmmathbf{k}+\tmmathbf{q}}} \frac{(q^0)^2 - [(\bm{k}\cdot\bm{q})/m]^2}{(q^0)^2 - (E_{\tmmathbf{k}} +
		E_{\tmmathbf{k}+\tmmathbf{q}})^2} = 0.
\end{equation}
Averaging this equation over the angle between $\bm{k}$ and $\bm{q}$~\cite{Littlewood1982}, we obtain
\begin{equation}
	q^0 = \pm v_F q / \sqrt{3},
\end{equation}
where $v_F:=k_F/m$ is the Fermi velocity. Here we note that the Fermi velocity decays in time since the particle loss leads to the shrinking of the Fermi surface. 

We now consider the correction to the NG mode due to dissipation. By assuming the dispersion relation in the form of $q^0 = v_F q / \sqrt{3} + i \gamma n f(q)$, we expand Eq. \eqref{eq:corrected} in terms of $\gamma$ and obtain
\begin{equation}\label{eq: first-order}
	- \frac{i \gamma}{2 U} \tau_2 + i \gamma n U \int \frac{k^2 d k}{2 \pi^2}
	\frac{v_F q f (q)}{4 \sqrt{3} \Delta^3} = i \int \frac{d^4 k}{(2 \pi)^4} U
	\tau_3 (G_1^T (k + q) \tau_2 G_0^T (k) + G_0^T (k + q) \tau_2 G_1^T (k))
	\tau_3,
\end{equation}
where
\begin{equation}
	G_0^T (k) = \frac{1}{(k^0)^2 - E_{\tmmathbf{k}}^2} (k^0 +
	\varepsilon_{\tmmathbf{k}} \tau_3 + \Delta \tau_1), G_1^T (k) = \frac{i
		\gamma n}{2 ((k^0)^2 - E_{\tmmathbf{k}}^2)} - \frac{i \gamma n
		\omega}{((k^0)^2 - E_{\tmmathbf{k}}^2)^2} (k^0 + \varepsilon_{\tmmathbf{k}}
	\tau_3 + \Delta \tau_1) .
\end{equation}
From Eq. \eqref{eq: first-order} we find that $f(q)$ appears in the terms proportional to $q^n$ with $n\geq1$. By considering the higher-order terms in $q$ and performing the integration over the frequency $k^0$, we have
\begin{equation}\label{eq: 1st-order}
	i \gamma n U \int \frac{k^2 d k}{2 \pi^2}
	\frac{v_F q f (q)}{4 \sqrt{3} \Delta^3} = i U \gamma n \int \frac{d^3
		k}{(2 \pi)^3} \frac{A_{1\tmmathbf{k}} + A_{2\tmmathbf{k}} +
		A_{3\tmmathbf{k}} + A_{4\tmmathbf{k}}}{4 E_{\tmmathbf{k}}^3 (q^0 +
		E_{\tmmathbf{k}} - E_{\tmmathbf{k}+\tmmathbf{q}})^2 (q^0 + E_{\tmmathbf{k}}
		+ E_{\tmmathbf{k}+\tmmathbf{q}})^2},
\end{equation}
where
\begin{eqnarray*}
	A_{1\tmmathbf{k}} & = & E_{\tmmathbf{k}}^2 (E_{\tmmathbf{k}+\tmmathbf{q}}^2
	\varepsilon_{\tmmathbf{k}+\tmmathbf{q}} + (3 \varepsilon_{\tmmathbf{k}} - 2
	\varepsilon_{\tmmathbf{k}+\tmmathbf{q}}) (\Delta^2 +
	\varepsilon_{\tmmathbf{k}} \varepsilon_{\tmmathbf{k}+\tmmathbf{q}}) -
	(q^0)^2 (\varepsilon_{\tmmathbf{k}+\tmmathbf{q}} + 2
	\varepsilon_{\tmmathbf{k}})),\\
	A_{2\tmmathbf{k}} & = & 4 q^0 E_{\tmmathbf{k}} \varepsilon_{\tmmathbf{k}}
	(\Delta^2 + \varepsilon_{\tmmathbf{k}}
	\varepsilon_{\tmmathbf{k}+\tmmathbf{q}}),\\
	A_{3\tmmathbf{k}} & = & - \varepsilon_{\tmmathbf{k}}
	(E_{\tmmathbf{k}+\tmmathbf{q}}^2 - (q^0)^2) (\Delta^2 +
	\varepsilon_{\tmmathbf{k}} \varepsilon_{\tmmathbf{k}+\tmmathbf{q}}),\\
	A_{4\tmmathbf{k}} & = & E_{\tmmathbf{k}}^4
	(\varepsilon_{\tmmathbf{k}+\tmmathbf{q}} - 2 \varepsilon_{\tmmathbf{k}}) - 4
	q^0 E_{\tmmathbf{k}}^3 \varepsilon_{\tmmathbf{k}} .
\end{eqnarray*}
By substituting $E_{\tmmathbf{k}} = \sqrt{\varepsilon_{\tmmathbf{k}}^2 + \Delta^2}$, we can rewrite the right-hand side of Eq. \eqref{eq: 1st-order} as
\begin{equation}
	i U \gamma n \int \frac{d^3 k}{(2 \pi)^3} \frac{- (q_0^2 (\Delta^2
		(\varepsilon_{\tmmathbf{k}} + \varepsilon_{\tmmathbf{k}+\tmmathbf{q}}) + 2
		\varepsilon_{\tmmathbf{k}}^3)) + 4 q_0 \varepsilon_{\tmmathbf{k}}
		E_{\tmmathbf{k}} (\varepsilon_{\tmmathbf{k}+\tmmathbf{q}} -
		\varepsilon_{\tmmathbf{k}})^2 - (\varepsilon_{\tmmathbf{k}} -
		\varepsilon_{\tmmathbf{k}+\tmmathbf{q}})^2 (\Delta^2
		(\varepsilon_{\tmmathbf{k}} - \varepsilon_{\tmmathbf{k}+\tmmathbf{q}}) + 2
		\varepsilon_{\tmmathbf{k}}^3)}{4 E_{\tmmathbf{k}}^3 (q^0 + E_{\tmmathbf{k}}
		- E_{\tmmathbf{k}+\tmmathbf{q}})^2 (q^0 + E_{\tmmathbf{k}} +
		E_{\tmmathbf{k}+\tmmathbf{q}})^2} .
\end{equation}
By taking the average over the angle between $\bm{k}$ and $\bm{q}$, we finally arrive at the result
\begin{equation}
	i U \gamma n \int \frac{k^2 d k}{2 \pi^2} \left( \frac{3 (v_F q)^3}{32
		\Delta^5} + O (q^4) \right) .
\end{equation}
Comparing this result with Eq. \eqref{eq: 1st-order}, we find
\begin{equation}
	i \gamma n U \frac{v_F q f (q)}{4 \sqrt{3} \Delta^3} = i U \gamma n \frac{3
		(v_F q)^3}{32 \Delta^5} \Rightarrow f (q) = \frac{3 \sqrt{3} (v_F q)^2}{8
		\Delta^2} \propto q^2 .
\end{equation}
This indicates that the correction corresponds to the diffusive mode. One can also show that the sign of the linear term does not influence the final result.

\section{Weak U(1) Symmetry Breaking}
In this section, we consider the Lindbladian that breaks weak U(1) symmetry and reconsider the Ward-Takahashi identity and response to the gauge field. We perform strong U(1) transformation of the action and similarly obtain the Ward-Takahashi identity as
\begin{align}\label{eq: Ward-open5}
	&   \pm[\delta (x - x_1) \tau_3 \langle \Psi_{\pm} (x_1) \bar{\Psi}_{\pm}
	(x_2) \rangle - \delta (x - x_2) \langle \Psi_{\pm} (x_1) \bar{\Psi}_{\pm}
	(x_2) \rangle \tau_3] = \langle \Psi_{\pm} (x_1) \bar{\Psi}_{\pm} (x_2)
	\partial_{\mu} \bar{J}_d^{\mu} \rangle, \\ \label{eq: Ward-open6}
	& \pm[\delta (x - x_1) \tau_3 \langle \Psi_{\pm} (x_1) \bar{\Psi}_{\mp}
	(x_2) \rangle + \delta (x - x_2) \langle \Psi_{\pm} (x_1) \bar{\Psi}_{\mp}
	(x_2) \rangle \tau_3] = \langle \Psi_{\pm} (x_1) \bar{\Psi}_{\mp} (x_2)
	\partial_{\mu} \bar{J}_d^{\mu} \rangle,
\end{align}
where the full current is given by
\begin{equation}
	\bar{J}_d^{\mu}:=J_c^{\mu}+J_{\rm{wd}}^{\mu}.
\end{equation}
Here $J_c^{\mu}$ is defined in Eq. \eqref{eq: bar_Jc} and $J_{\rm{wd}}^{\mu}$ is defined as
\begin{align}
	\nabla\cdot \bm{J}_{\rm{wd}} = i\gamma \left(
	\bar{L}_{\tmmathbf{r}-} \frac{\partial L_{\tmmathbf{r}+}}{\partial
		\theta(x)} + \frac{\partial \bar{L}_{\tmmathbf{r}-}}{\partial
		\theta(x)} L_{\tmmathbf{r}+} \right)-i\frac{\gamma}{2}\left(
	\frac{\partial(\bar{L}_{\tmmathbf{r}+}\partial L_{\tmmathbf{r}+})}{\partial
		\theta(x)} +  \frac{\partial(\bar{L}_{\tmmathbf{r}-}\partial L_{\tmmathbf{r}-})}{\partial
		\theta(x)} \right),\ J_{\rm{wd}}^0=0.
\end{align}
Compared with Eq. \eqref{eq: jc}, there emerges an additional term
\begin{equation}
	\nabla\cdot\bm{J}_{\text{da}}:=-i\frac{\gamma}{2}\left(
	\frac{\partial(\bar{L}_{\tmmathbf{r}+}\partial L_{\tmmathbf{r}+})}{\partial
		\theta(x)} +  \frac{\partial(\bar{L}_{\tmmathbf{r}-}\partial L_{\tmmathbf{r}-})}{\partial
		\theta(x)} \right).
\end{equation}
In this case, if we perform the gauge transformation $A_{\mu}\to A_{\mu}+\partial_{\mu}\phi$, the perturbed current takes the form of
\begin{align}
	\delta J_2^{\mu} (x) & = - \frac{i}{2}\sum_{\alpha} \int d x' \Tr \langle
	\gamma_{\alpha}^{\mu} \Psi_{\alpha} (x) J_{\text{da}}^{\nu} (x') \bar{\Psi}_{\alpha} (x)
	\rangle \partial_{\nu}' \phi (x') \nonumber\\
	&= \frac{i}{2}\sum_{\alpha} \int d x' \Tr \langle
	\gamma_{\alpha}^{\mu} \Psi_{\alpha} (x) \partial_{\nu}' J_{\text{da}}^{\nu} (x') \bar{\Psi}_{\alpha} (x)
	\rangle\phi (x')\nonumber\\
	&=\frac{\gamma}{4}\sum_{\alpha,\alpha'} \int d x' \Tr \left\langle
	\gamma_{\alpha}^{\mu} \Psi_{\alpha} (x)  \frac{\partial(\bar{L}_{\tmmathbf{r}'\alpha'}\partial L_{\tmmathbf{r}'\alpha'})}{\partial
		\theta(x')} \bar{\Psi}_{\alpha} (x)
	\right\rangle\phi (x'),
\end{align}
which breaks the gauge invariance. Therefore, the minimal condition for the gauge-invariant transport theory is weak U(1) symmetric Lindbladian.

Finally, we address the issue of an experimental signature of the weak U(1) symmetry of the Lindbladian since $J^\mu_q:=J^\mu_+-J^\mu_-$ is not an observable. We here focus on the dynamics of a quantity
\begin{align}\label{eq:OTOC}
	O_N(t):=\Tr[N\rho(t)N\rho(t)]-\Tr[N^2\rho(t)^2]=-\frac{1}{2}\langle\langle\rho|(N_{+}-N_{-})^2|\rho\rangle\rangle,
\end{align} 
where $|\rho\rangle\rangle$ is the vectorized density matrix defined by $|\rho\rangle\rangle:=\sum_{i,j}\rho_{ij}|i\rangle|j\rangle$ for the density matrix $\rho=\sum_{i,j}\rho_{ij}|i\rangle\langle j|$. If the Lindbladian and the initial density matrix satisfy the weak U(1) symmetry, the quantity $O_N(t)$ remains invariant in the dynamics. The conservation of the observable $O_N(t)$ is associated with the equation of continuity of the weak U(1) symmetry since it depends only on $N_+-N_-$. In the main text, we have directly shown this point. 

Let us now consider the dynamics of an observable $O_N(t)$ in dissipative BCS theory~\cite{Yamamoto2021}, which is described by the action \eqref{eq: BCS_action}. Since there is no off-diagonal terms in the contour basis, it is consistent to use the time-dependent BCS ansatz:
\begin{equation}
	|\Psi_{\text{BCS}}(t)\ket=\prod_{\bm{k}}(u_{\bm{k}}(t)+v_{\bm{k}}(t)c_{\bm{k}\uparrow}^{\dagger}c_{-\bm{k}\downarrow}^{\dagger})|0\ket,
\end{equation} 
where the initial condition is given by
\begin{equation}
	u_{\bm{k}}(0)=\sqrt{\frac{E_{\bm{k}}+\varepsilon_{\bm{k}}}{2E_{\bm{k}}}},\ v_{\bm{k}}(0)=\sqrt{\frac{E_{\bm{k}}-\varepsilon_{\bm{k}}}{2E_{\bm{k}}}},\ E_{\bm{k}}=\sqrt{\varepsilon_{\bm{k}}^2+|\Delta|^2},
\end{equation}
and the order parameter is
\begin{equation}
	\Delta=-\frac{U_c}{N}\sum_{\bm{k}}\bra c_{-\bm{k}\downarrow}c_{\bm{k}\uparrow}\ket=-\frac{U_c}{N}\sum_{\bm{k}}u_{\bm{k}}^{*}(t)v_{\bm{k}}(t).
\end{equation}
Recall that $U_c:=U+i\gamma/2$ is complex-valued. The dynamics of $u_{\bm{k}}$ and $v_{\bm{k}}$ is governed by
\begin{equation}
	i\partial_t\left(\begin{array}{c}
		u_{\bm{k}}(t)\\
		v_{\bm{k}}(t)
	\end{array}\right)=\left(\begin{array}{cc}
		-\varepsilon_{\bm{k}} & \Delta^{\ast}\\
		\Delta & \varepsilon_{\tmmathbf{k}}
	\end{array}\right)\left(\begin{array}{c}
		u_{\bm{k}}(t)\\
		v_{\bm{k}}(t)
	\end{array}\right),
\end{equation}
which preserves $|u_{\bm{k}}|^2+|v_{\bm{k}}|^2=1$. We can simplify the observable $O_N(t)$ defined in Eq. \eqref{eq:OTOC} as
\begin{align}\label{eq: ON}
	O_N(t)&=|\bra\Psi_{\text{BCS}}|N|\Psi_{\text{BCS}}\ket|^2-\bra\Psi_{\text{BCS}}|N^2|\Psi_{\text{BCS}}\ket\nonumber\\
	&=\left(2\sum_{\bm{k}}|v_{\bm{k}}|^2\right)^2-\left(4\sum_{\bm{k},\bm{k}'(\bm{k}'\neq\bm{k})}|v_{\bm{k}}|^2|v_{\bm{k}'}|^2+4\sum_{\bm{k}}|v_{\bm{k}}|^2\right)\nonumber\\
	&=4\sum_{\bm{k}}|v_{\bm{k}}|^4-4\sum_{\bm{k}}|v_{\bm{k}}|^2\nonumber\\
	&=-2N(t)+\sum_{\bm{k}}n_{\bm{k}}^2,
\end{align}
where we use the relation $N(t)=\sum_{\bm{k}}n_{\bm{k}}=2\sum_{\bm{k}}|v_{\bm{k}}|^2$ with $n_{\bm{k}}=2|v_{\bm{k}}|^2$ being the number of particles with a momentum $\bm{k}$ and 
\begin{align}
	\bra\Psi_{\text{BCS}}|N^2|\Psi_{\text{BCS}}\ket&=\bra\Psi_{\text{BCS}}|N\sum_{\bm{k}}\left[2v_{\bm{k}}c_{\bm{k}\uparrow}^{\dagger}c_{-\bm{k}\downarrow}^{\dagger}\prod_{\bm{k}'\neq\bm{k}}(u_{\bm{k}'}+v_{\bm{k}'}c_{\bm{k}'\uparrow}^{\dagger}c_{-\bm{k}'\downarrow}^{\dagger})\right]|0\ket\nonumber\\
	&=4\sum_{\bm{k}}|v_{\bm{k}}|^2+\bra\Psi_{\text{BCS}}|\sum_{\bm{k},\bm{k}'(\bm{k}'\neq\bm{k})}\left[(2v_{\bm{k}}c_{\bm{k}\uparrow}^{\dagger}c_{-\bm{k}\downarrow}^{\dagger})(2v_{\bm{k}'}c_{\bm{k}'\uparrow}^{\dagger}c_{-\bm{k}'\downarrow}^{\dagger})\prod_{\bm{k}''\neq\bm{k},\bm{k}'}(u_{\bm{k}''}+v_{\bm{k}''}c_{\bm{k}''\uparrow}^{\dagger}c_{-\bm{k}''\downarrow}^{\dagger})\right]|0\ket\nonumber\\
	&=2N(t)+4\sum_{\bm{k},\bm{k}'(\bm{k}'\neq\bm{k})}|v_{\bm{k}}|^2|v_{\bm{k}'}|^2.
\end{align}
By using the relation $|u_{\bm{k}}|^2+|v_{\bm{k}}|^2=1$, we can rewrite Eq. \eqref{eq: ON} as
\begin{equation}
	O_N(t)=-4\sum_{\bm{k}}|u_{\bm{k}}|^2|v_{\bm{k}}|^2,
\end{equation}
which remains negative until all the particles are lost into the environment. This result indicates that the mean-field theory fails in predicting the dynamics of the quantity $O_N(t)$ and affirms the importance of preserving the gauge invariance and the weak U(1) symmetry.

\end{document}